\documentclass[sigplan, table, screen]{acmart}

\copyrightyear{2020}
\acmYear{2020}
\setcopyright{acmlicensed}
\acmConference[ASPLOS '20]{Proceedings of the Twenty-Fifth International Conference on Architectural Support for Programming Languages and Operating Systems}{March 16--20, 2020}{Lausanne, Switzerland}
\acmBooktitle{Proceedings of the Twenty-Fifth International Conference on Architectural Support for Programming Languages and Operating Systems (ASPLOS '20), March 16--20, 2020, Lausanne, Switzerland}
\acmPrice{15.00}
\acmDOI{10.1145/3373376.3378475}
\acmISBN{978-1-4503-7102-5/20/03}

\usepackage[inline]{enumitem}
\usepackage{tikz}
\usetikzlibrary{shapes.misc,arrows.meta}
\usepackage[noend]{algpseudocode}
\usepackage{algorithm}
\usepackage{multicol}
\usepackage{bbm}
\usepackage{cleveref}
\usepackage{dashrule}
\usepackage{flushend}

\usepackage{adjustbox}
\newcommand\VRule[1][\arrayrulewidth]{\vrule width #1}
\usepackage{pifont}
\newcommand{\cmark}{\ding{51}}%
\newcommand{\xmark}{\ding{55}}%
\newcolumntype{C}{>{\centering\arraybackslash}c}

\include{macros}

\begin{document}

\title{Atomicity Checking in Linear Time using Vector Clocks}

\author{Umang Mathur}
\email{umathur3@illinois.edu}
\orcid{0000-0002-7610-0660}
\affiliation{
  \institution{University of Illinois at Urbana-Champaign}
}

\author{Mahesh Viswanathan}
\email{vmahesh@illinois.com}
\affiliation{%
  \institution{University of Illinois at Urbana-Champaign}
}


\begin{abstract}
Multi-threaded programs are challenging to write. Developers often
need to reason about a prohibitively large number of thread
interleavings to reason about the behavior of software.  A
non-interference property like atomicity can reduce this interleaving
space by ensuring that any execution is equivalent to an execution
where all atomic blocks are executed serially.  We consider the well
studied notion of conflict serializability for dynamically checking
atomicity.  Existing algorithms detect violations of conflict
serializability by detecting cycles in a graph of transactions
observed in a given execution.  
The number of edges in such a graph can grow quadratically with the length of the trace making the analysis not scalable.  
In this paper, we present~\algo, a novel single pass linear time
algorithm that uses vector clocks to detect violations of 
conflict serializability in an online setting.  
Experiments show that {\algo} scales to traces with a
large number of events with significant speedup.
\end{abstract}

\ccsdesc[500]{Software and its engineering~Dynamic analysis}
\ccsdesc[500]{Software and its engineering~Software testing and debugging}

\keywords{Concurrency, Atomicity, Conflict Serializability, Vector Clocks, Dynamic Program Analysis}

\maketitle

\section{Introduction}
\seclabel{intro}

Writing correct multi-threaded programs is extremely difficult. It is
the class of software that is most prone to errors. Reasoning about
multi-threaded programs is notoriously challenging due to the
inherent nondeterminism that arises from thread scheduling in such
systems. If the program satisfies certain fundamental properties then
reasoning about them becomes easier, and if such properties are
violated then it is often symptomatic of more serious bugs in the
software. \emph{Atomicity} is one such classical concurrency property,
which guarantees that a programmer reasoning about a concurrent
program can assume that atomic blocks of code can be executed
sequentially without any context switches in between. Atomicity allows
programmers to reason about atomic blocks without worrying about the
effects of other threads. Unfortunately, violation of atomicity specifications
is quite common and is the root cause in a majority of real-world
bugs~\cite{lpsz08,ff04,TaxDC2016,Wang2017,Liu2017,Fonseca2017,Chew2010}.

Various approaches to identifying atomicity violations have been
explored. Static analysis based approaches for atomicity checking are
usually conservative, computationally expensive, and often rely on
user annotations, like type
annotations~\cite{fq03,fq03-2,as04,ffq05,saws05,ws05}. The advantage
of static analysis approaches is that they may successfully
\emph{prove} that a program satisfies all its atomicity
requirements. Dynamic analysis for atomicity violations, on the other
hand, have the advantage that they are fully automated and are
computationally less
expensive~\cite{ff04,xbh05,ltqz06,fm08,velodrome,doublechecker}. Though
they cannot prove that a program satisfies its atomicity
specification, dynamic analysis can be used to check if an observed
trace is witness to the violation of atomicity. Given their
scalability, dynamic analysis techniques for detecting 
atomicity violations have proved to be very useful in practice.

In this paper, we will focus on sound and precise dynamic analyses;
unsound dynamic analyses have the disadvantage that they report many
false alarms\footnote{We use the term \emph{sound} for a dynamic
  analysis technique if it does not report false alarms. This is
  consistent with the usage of the term ``sound'' in the context of dynamic
  analyses~\cite{soundness-dynamic-analysis}.}.
Most sound and precise dynamic analyses~\cite{fm08,velodrome,doublechecker}
for atomicity violation are based on checking the \emph{conflict
  serializability} of an observed program execution. An execution is
conflict serializable if it can be transformed into an equivalent
execution, where all statements in an atomic block are executed
consecutively without context switches, by commuting adjacent,
non-conflicting operations of different threads. Here conflicting
operations are either two operations by the same thread, two accesses
(at least one of which is a write access) to a common memory location, or acquires and
releases of common locks. Determining if an execution is conflict
serializable can be reduced to checking for the existence of a cycle
in a graph called the transaction graph. The transaction graph has
atomic blocks (a.k.a. transactions) as vertices, and edges between
blocks that contain non-commutable events. A path from atomic block
$A$ to $B$ indicates that $A$ must be executed before $B$ in a serial
execution, and so a cycle in such a graph indicates that the execution
is not equivalent to a serial one. 
All current sound and precise dynamic analyses for conflict
serializability~\cite{velodrome,doublechecker} 
rely on this idea
and thus have an asymptotic complexity of cubic time --- each new event
of the trace requires updating the transaction graph, and checking for
cycles; the number of edges can be quadratic 
in the number of events, giving a quadratic processing time \emph{per event}.

The central question motivating this paper is the following: Is a
cubic running time necessary for checking conflict
serializability? Or are there sub-cubic algorithms for this
problem? The main result of this paper is a new, \emph{linear time} algorithm
for checking conflict serializability.

For other concurrency specifications, like data race detection, that
admit sound and precise linear time  algorithms, the key to achieving
an efficient algorithm is the use of vector
clocks~\cite{Mattern1988,wcp2017,shb2018,genc19}. Such algorithms rely on computing vector
timestamps for events in a streaming fashion as the trace is
generated, and using these timestamps to recover the causal order
between a pair of events. However, generalizing such an algorithmic
principle to conflict serializability checking is far from
straightforward. This is because checking conflict serializability
requires identifying causal orders between transactions (or atomic
blocks) and not individual events. For this reason,
Flanagan-Freund-Yi~\cite{velodrome}, in fact, dismiss the possibility
of a vector clock based algorithm for conflict serializability
checking:
\begin{quote}
``The traditional representation of clock vectors~\cite{Mattern1988}
is not applicable because our happens-before relation is over compound
transactions and not individual operations.''
\end{quote}
The challenge is to discover a way to associate a single timestamp
with a transaction, even though new causal dependencies are discovered
as each individual event in the trace is processed. This is further
complicated by the following observation. Vector timestamps implicitly
summarize the set of all events that must be ordered before. However,
the set of transactions that must be executed before a transaction $T$
might be known only well after all the events of $T$ have been seen
(see \exref{pred}). These observations suggest that a scheme of
assigning vector timestamps to transactions may only be computed if
the algorithm makes multiple streaming passes over the trace, 
which may result in an algorithm that is not linear time.

We address these challenges by assigning vector timestamps to
individual events in a trace. The induced order on events is then used
to discover the ordering relationship between transactions, and
thereby determining if a trace is conflict serializable. For a trace
containing a bounded number of variables, threads, and locks, our
algorithm, \algo, is a single pass, streaming algorithm that runs in linear
time~\footnote{Vector clock based algorithms are linear time under the
computational assumption that arithmetic operations take constant
time. This is a reasonable assumption because even for
traces with billions of events, the numbers involved in vector clocks
can be stored in a single word, and so addition and subtraction of
such numbers can be reasoned to be in constant time.}.
As with standard vector clock algorithms, such as those 
used in data race detection~\cite{Pozniansky:2003:EOD:966049.781529,fasttrack},
our algorithm summarizes information in \emph{vector clocks}
and thus does not need to store the timestamp of all the events
in the trace to detect serializability violations.

We have implemented \algo~in our tool 
{\tool}~\cite{rapid} and have compared its performance against
Velodrome~\cite{velodrome} on various benchmark programs. 
Atomicity specifications (i.e., which blocks of code should be
regarded as atomic) are hard to come by. One na\"{i}ve specification
is to consider each method call to be atomic. Since often there is a
\texttt{main} method for each thread, this means that the entire
computation of each thread should be atomic. Programs are unlikely to
satisfy such strong atomicity specifications, but running detection
algorithms against these, gives us a baseline. We use such na\"{i}ve
specifications for some programs in our benchmark. 
For such benchmarks, conflict serializability is trivially violated 
in a small prefix of the observed trace.
The resulting transaction graph is thus small,
the overhead of  maintaining vector clocks outweighs the benefits of a linear time
algorithm, and Velodrome slightly outperforms~\algo. For
other programs in our benchmark, we use the more realistic atomicity
specifications given in~\cite{doublechecker}. Here transactions
consist of smaller blocks of code, and the resulting transaction graph
has many transactions. For such examples, our algorithm significantly
outperforms Velodrome. This suggests that on realistic atomicity
specifications, the benefits of having a linear time algorithm can be
significant.

The rest of the paper is organized as follows.  In~\secref{prelim}, we
discuss preliminary notations such as that of concurrent program
traces and the definition of conflict serializability.
In~\secref{challenges}, we use motivating examples to illustrate the
challenges involved in developing a linear time vector clock algorithm
for dynamically checking conflict serializability.  In~\secref{vc}, we
discuss~\algo, a single pass linear time vector clock algorithm for
checking conflict serializability, which is also the main contribution
of the paper.  \secref{vc}~also discusses the correctness and
complexity guarantees of the algorithm and optimizations for improving
the performance of~\algo.  Our implementation of~\algo~in our
tool~\tool~and its performance evaluation on a suite of benchmark
programs is discussed in~\secref{experiments}.  We discuss closely
related work in~\secref{related} and present concluding remarks
in~\secref{conclusions}.  Some proofs and additional discussion can be
found in the full version~\cite{aerodrome-tech}.


\section{Preliminaries}
\seclabel{prelim}


%
%
An execution trace (or simply \emph{trace}) of a 
concurrent program is a sequence of events.
We will use $\tr, \rho_1,
\rho_2, \ldots$ to denote traces.  
Each event in a trace is a pair $e = \ev{t,
  op}$, where $t$ denotes the thread that performs $e$ and $op$ is the
operation performed by $e$; we will use $\thread(e)$ to denote $t$ and
$\op(e)$ to denote $op$.
Operations can be one of $\rd(x), \wt(x)$ (read
from or write to variable/memory location $x$), $\acq(\lk)$, $\rel(\lk)$
(acquire or release of lock object $\lk$), $\fork(u)$, $\join(u)$
(fork or join of thread $u$), $\trbegin$ or $\trend$ (denoting the
begin or end of an atomic block). 
Traces are assumed to be well-formed --- all lock acquires and releases
are well matched, a lock is not acquired by more than one thread at a time,
all begin and end events are well matched,
fork events occur before the first event of the child thread and
join events occur after the last event of the child thread.
A \emph{transaction} $T$ in thread $t$ is a maximal
subsequence~\footnote{We allow for nested blocks of begins and
  ends. In this case only the outermost begin and end constitute a
  transaction.} of events of thread $t$ that starts with $\ev{t,
  \trbegin}$ and ends with the matching $\ev{t, \trend}$, and we say
$e \in T$ if the event $e$ belongs to this maximal subsequence; in
this case, $\trnsc(e)$ denotes the transaction $T$ to which $e$
belongs. In a trace $\tr$, we will say that a transaction $T$ is
\emph{completed} in $\tr$ if the corresponding end transaction event
$\ev{\cdot,\trend} \in \tr$. If $T$ is not completed in $\tr$, it is
said to be \emph{active}. 


%

Given a trace $\tr$, we denote by $\trord{\tr}$ the total order on
events induced by $\tr$ --- for events $e, e'$ in $\tr$, we say
$e \trord{\tr} e'$ iff either $e = e'$ or $e$ occurs before
$e'$ in the sequence $\tr$.  Two events $e, e'$ are said to be
\emph{conflicting} if either
\begin{enumerate*}[label=(\roman*)]
	\item $\thread(e) = \thread(e')$,
	\item $e = \ev{t, \fork(u)}$ and $\thread(e') = u$,
	\item $\thread(e) = u$ and $e' = \ev{t, \join(u)}$,
	\item there is a common memory location $x$ such that
	both $\op(e), \op(e')$ are one of $\set{\wt(x), \rd(x)}$ and
	not both are $\rd(x)$, or
	\item there is a lock $\lk$ such that $\op(e) = \rel(\lk)$ and
	$\op(e') = \acq(\lk)$.
\end{enumerate*} 
Given a trace $\tr$, \emph{conflict-happens-before} $\chb{\tr}$
is the smallest reflexive, transitive relation such that
for every pair of conflicting events $e \trord{\tr} e'$, we have
$e \chb{\tr} e'$.

%
%
%
Atomicity is closely related to the property of conflict
serializability.
Informally, this property requires that an execution be
equivalent to a \emph{serial} execution by commuting adjacent non-conflicting
events; an execution is serial if for every
thread $t$ in the trace and for every transaction $T$
of thread $t$, there are no events of any other thread
between the begin and end events of $T$. 
In this context, if two events $e$ and $e'$ are
ordered by $\chb{}$, then their order is the same in all equivalent
executions. To capture conflict serializability, such a causal
relationship needs to be lifted to transactions. Consider two
transactions $T$ and $T'$ with events $e \in T$ and $e' \in
T'$ such that $e \chb{\tr} e'$. If the goal in a serial execution
is to schedule all events of $T$ consecutively, given that $e$ is
before $e'$ in all equivalent executions, it must be the case that
\emph{every} event of $T$ should happen before each event of $T'$. Thus,
transaction $T$ must \emph{happen before} transaction $T'$ in trace
$\tr$ (denoted $T \thb{\tr} T'$) if there are events $e \in T$
and $e' \in T'$ such that $e \chb{\tr} e'$. We now present the
definition of conflict serializability (which implies atomicity)
from~\cite{velodrome}.
\begin{definition}[Conflict Serializability~\cite{velodrome}]
\deflabel{conflict-serializability} 
A trace $\tr$ is \emph{conflict serializable} if there is no sequence
of $k>1$ distinct transactions $T_0, T_1 \ldots T_{k-1}$ such that 
for every $0 \leq i \leq {k{-}1}$, we have
$T_i \thb{\tr} T_{(i+1)\bmod k}$.
If $\tr$ is not conflict serializable, then such a 
sequence $T_0, \ldots, T_{k-1}$ is said to be a witness to the violation.
\end{definition}


\begin{figure}[t]
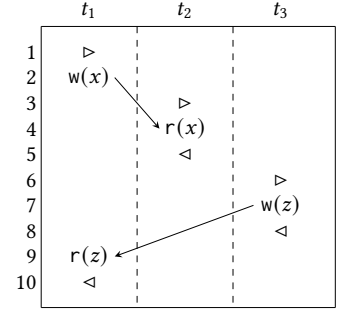

\centering
\execution{3}{
  \figev{1}{\trbeginfig}
  \figev{1}{\wt(x)}
  \figev{2}{\trbeginfig}
  \figev{2}{\rd(x)}
  \figev{2}{\trendfig}
  \figev{3}{\trbeginfig}
  \figev{3}{\wt(z)}
  \figev{3}{\trendfig}
  \figev{1}{\rd(z)}
  \figev{1}{\trendfig}
  \orderedge{1}{2}{0.4}{2}{4}{-0.4}
  \orderedge{3}{7}{-0.43}{1}{9}{0.4}
}
\caption{Trace $\rho_1$. Taking $T_i$ to be the transaction of thread $t_i$, we have $T_3 \thb{\rho_1} T_1 \thb{\rho_1} T_2$.}
\Description[Trace rho-1 with three threads]{The figure shows a trace, named rho-1, having three threads t-1, t-2 and t-3 performing 10 events in total. 
Event-1 is a begin-transaction event by t-1.
Event-2 is a write event to memory location x by t-1.
Event-3 is a begin-transaction event by t-2.
Event-4 is a read event from memory location x by t-2.
Event-5 is an end-transaction event by t-2.
Event-6 is a begin-transaction event by t-3.
Event-7 is a write event to memory location z by t-3.
Event-8 is an end-transaction event by t-3.
Event-9 is a read event from memory location z by t-1.
Event-10 is an end-transaction event by t-1.
There is a cross-thread CHB ordering from event-2 (which is a write to memory location x in thread t-1) to event-4 (which is a read from memory location x in thread t-2).
There is a cross-thread CHB ordering from event-7 (which is a write to memory location z in thread t-3) to event-9 (which is a read from memory location z in thread t-1).
}
\figlabel{trace-pred-change}
\end{figure}

\begin{example}
Consider the trace $\rho_1$ in~\figref{trace-pred-change}.
This trace is a sequence of 10 events, performed by three different
threads $t_1, t_2$ and $t_3$.
In all our examples, we will use
$e_i$ to denote the $i^\text{th}$ event in the trace.
This trace has three transactions --- 
transaction $T_1 = e_1e_2e_9e_{10}$ is performed in $t_1$,
transaction $T_2 = e_3e_4e_5$ is performed in $t_2$
and transaction $T_3 = e_6e_7e_8$ is performed in $t_3$. 
All pairs of events, both of which are
performed by the same thread (such as $(e_1, e_2)$ or $(e_2, e_{10})$
in $\rho_1$) are conflicting.
In addition, $(e_2, e_4)$ and $(e_7, e_9)$ 
are conflicting pairs of events in $\rho_1$
and we use an explicit arrow ({\protect\drawarrow})
to depict such inter-thread conflicting pairs.
We have $T_1 \thb{\rho_1} T_2$
because $e_2 \chb{\rho_1} e_4$ and $T_3 \thb{\rho_1} T_1$
because $e_7 \chb{\rho_1} e_9$.
Also note that $\chb{}$ is a transitive order and thus
$e_1 \chb{\rho_1} e_5$ because 
$e_1 \chb{\rho_1} e_2$, $e_2 \chb{\rho_1} e_4$ and $e_4 \chb{\rho_1} e_5$.
Finally, the trace $\rho_1$ is conflict serializable and
the equivalent serial execution is the sequence
$\rho^\text{serial}_1 = e_6e_7e_8e_1e_2e_9e_{10}e_3e_4e_5$,
in which the order of transaction is $T_3T_1T_2$.
Observe that the relative
order of conflicting events in $\rho^\text{serial}_1$
is the same as in the original trace $\rho_1$.
\end{example}

Based on \defref{conflict-serializability}, a cyclic
dependency on transactions using $\thb{\tr}$ suggests that $\tr$ does
not have an equivalent serial execution and hence the program does not
satisfy its atomicity specification. Previous
techniques~\cite{velodrome,doublechecker} for checking conflict
serializability dynamically, rely on constructing a directed graph.
The vertices in such a graph are the different transactions in the
observed trace, the edges correspond to the order imposed by $\thb{}$
and checking violations of conflict serializability reduces to searching for
a cycle in this graph. These algorithms run in time that is cubic in
the length of the observed trace as they check for cycles
each time a new edge is added in the graph, whose size is quadratic in the size of the trace.



\section{Challenges in Designing a Vector Clock Algorithm}
\seclabel{challenges}

Vector clocks have been very useful in designing linear time
algorithms for dynamic analysis of multi-threaded
systems~\cite{djit,Pozniansky:2003:EOD:966049.781529,fasttrack,wcp2017,shb2018,genc19,Roemer18}. The
broad principle behind these algorithms, is to assign 
vector \emph{timestamps} 
to events as the trace is generated/observed so
that the ordering between these assigned timestamps captures causal ordering. 
Notice that, conflict serializability is defined in terms of the relation $\thb{}$
on transactions (\defref{conflict-serializability}),
and thus, the most
straightforward vector clock algorithm would rely on assigning
timestamps to \emph{transactions} in such a way that the timestamp of 
transaction $T_1$
is less than or equal to timestamp of transaction $T_2$ if and only if $T_1 \thb{}
T_2$. 
However, since a transaction is a \emph{sequence of events} (and not a
single event), the first challenge is figuring out how to assign and update
timestamps of transactions when individual events are being continuously
generated by the execution; this is one of the
reasons why such algorithms were deemed impossible for atomicity
in~\cite{velodrome}. However, there is a deeper and more fundamental
challenge with assigning timestamps to transactions, as illustrated in
the following example.

\begin{example}
\exlabel{pred} 
Consider again the trace $\rho_1$ in~\figref{trace-pred-change}.  
%
Notice that there is a ``path'' from $T_3$ to $T_2$ (via $T_1$) using
$\thb{\rho_1}$, even though $T_3$ starts after $T_2$  is completed in the trace
$\rho_1$. Further the discovery that $T_3$ has a path to $T_2$ can be
made only after the event $e_9$ is generated in the trace, and
at that point, \emph{both} $T_2$ and $T_3$ have completed. This poses
serious challenges when designing a vector clock algorithm. A vector
clock algorithm assigning a timestamp to transaction $T$ that is
consistent with $\thb{}$, needs to know (explicitly or implicitly) the
set of transactions that have a path to $T$; this is because the
algorithm needs to ensure that the timestamp assigned to $T$ is
ordered after
the timestamps assigned to all these ``predecessor''
transactions. However, as transaction $T_2$ in trace $\rho_1$
illustrates, this may require knowing future events and transactions.
\end{example}

\exref{pred} illustrates that transactions $T'$ that have a
$\thb{}$-path to a transaction $T$ may only be determined by events
that appear after $T$ itself. This suggests that one is unlikely to
get a linear time streaming algorithm that assigns timestamps
to transactions for detecting atomicity violations.

Therefore, we explore the possibility of an algorithm that assigns
timestamps to \emph{events} (not transactions), but which can
nonetheless enable checking conflict serializability. The first key
question to address is \emph{which relation among events should the
timestamps try to capture implicitly?} 
Recall that, the relation $\thb{}$ (on transactions) is defined
in terms of the relation $\chb{}$ (on events), 
and therefore, a natural first step to explore,
is to see if computing $\chb{}$ is sufficient to detect atomicity
violations.


\begin{figure}[h]
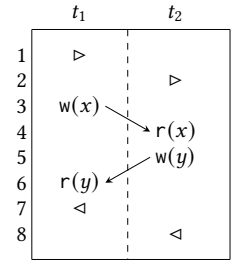

\centering
\execution{2}{
  \figev{1}{\trbeginfig}
  \figev{2}{\trbeginfig}
  \figev{1}{\wt(x)}
  \figev{2}{\rd(x)}
  \figev{2}{\wt(y)}
  \figev{1}{\rd(y)}
  \figev{1}{\trendfig}
  \figev{2}{\trendfig}
  \orderedge{1}{3}{0.4}{2}{4}{-0.4}
  \orderedge{2}{5}{-0.4}{1}{6}{0.4}
}
\caption{Trace $\rho_2$. There is a cycle in the transaction graph
that can be realized by a path using $\chb{}$-edges that begins and ends in the 
same transaction.}
\Description[Trace rho-2 with two threads]{The figure shows a trace, named rho-2, having two threads t-1 and t-2 performing 8 events in total. 
Event-1 is a begin-transaction event by t-1.
Event-2 is a begin-transaction event by t-2.
Event-3 is a write event to memory location x by t-1.
Event-4 is a read event from memory location x by t-2.
Event-5 is a write event to memory location y by t-2.
Event-6 is a read event from memory location y by t-1.
Event-7 is an end-transaction event by t-1.
Event-8 is an end-transaction event by t-2.
There is a cross-thread CHB ordering from event-3 (which is a write to memory location x in thread t-1) to event-4 (which is a read from memory location x in thread t-2).
There is a cross-thread CHB ordering from event-5 (which is a write to memory location y in thread t-2) to event-6 (which is a read from memory location y in thread t-1).
}
\figlabel{trace-timestamps-events}
\end{figure}

\begin{example}
\exlabel{timestamps-events}
Consider the trace $\rho_2$ in~\figref{trace-timestamps-events} with two
transactions $T_1$ and $T_2$ in threads $t_1$ and $t_2$ respectively.
Here, we have, 
$T_1 \thb{\rho_2} T_2$ and $T_2\thb{\rho_2} T_1$,
thus giving us a violation of conflict serializability
with the sequence $T_1, T_2$ witnessing the violation.
Now consider the following $\chb{}$ path
in the trace --- $e_1 \chb{\rho_2} e_4 \chb{\rho_2} e_5 \chb{\rho_2} e_7$.
This path, in fact, is symptomatic of the atomicity violation
because it starts and ends in the
same transaction (transaction $T_1$) 
and passes through another transaction (transaction $T_2$).
\end{example}

The atomicity violation in trace $\rho_2$ in \exref{timestamps-events}
can be deduced based on the observation that there are 3 events
$e,f,g$ ($e_1,e_5,e_7$ in $\rho_2$, specifically) such that $\trnsc(e)
= \trnsc(g)$, $\trnsc(e) \neq \trnsc(f)$, and $e \chb{} f \chb{}
g$. If we can prove that this is equivalent to
\defref{conflict-serializability}, then all we need to do is to
compute (implicitly using vector clocks) the $\chb{}$
ordering. Unfortunately, this is not true, i.e., violations of
conflict serializability cannot be detected by simply using
$\chb{}$ ordering and searching for the above kind of $\chb{}$ paths. 
We illustrate this in the next example.


\begin{figure}[h]
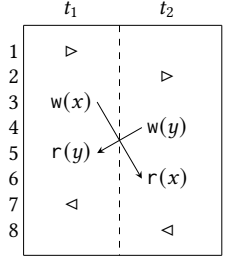

\centering
\execution{2}{
  \figev{1}{\trbeginfig}
  \figev{2}{\trbeginfig}
  \figev{1}{\wt(x)}
  \figev{2}{\wt(y)}
  \figev{1}{\rd(y)}
  \figev{2}{\rd(x)}
  \figev{1}{\trendfig}
  \figev{2}{\trendfig}
  \orderedge{1}{3}{0.4}{2}{6}{-0.4}
  \orderedge{2}{4}{-0.4}{1}{5}{0.4}
}
\caption{Trace $\rho_3$. There is no $\chb{}$ path that starts and ends in the same transaction.}
\Description[Trace rho-3 with two threads]{The figure shows a trace, named rho-3, having two threads t-1 and t-2 performing 8 events in total. 
Event-1 is a begin-transaction event by t-1.
Event-2 is a begin-transaction event by t-2.
Event-3 is a write event to memory location x by t-1.
Event-4 is a write event to memory location y by t-2.
Event-5 is a read event from memory location y by t-1.
Event-6 is a read event from memory location x by t-2.
Event-7 is an end-transaction event by t-1.
Event-8 is an end-transaction event by t-2.
There is a cross-thread CHB ordering from event-3 (which is a write to memory location x in thread t-1) to event-6 (which is a read from memory location x in thread t-2).
There is a cross-thread CHB ordering from event-4 (which is a write to memory location y in thread t-2) to event-5 (which is a read from memory location y in thread t-1).
}
\figlabel{trace-no-chb-path}
\end{figure}

\begin{example}
\exlabel{no-chb-path} 
Consider trace $\rho_3$ in~\figref{trace-no-chb-path}. 
As before, let $T_1$, $T_2$ be the two transactions by threads
$t_1$ and $t_2$ respectively.
Here, both $T_1 \thb{\rho_3} T_2$
(because $e_3  \chb{\rho_3} e_6$) 
and $T_2 \thb{\rho_3} T_1$ 
(because $e_4 \chb{\rho_3} e_5$),
thus giving us a conflict serializability violation. 
However, there is no
$\chb{}$-path that starts and ends in the same transaction. 
If vector
timestamps are used to compute $\chb{}$, then violations of conflict
serializability cannot be detected by checking ordering of vector
timestamps of events.
\end{example}

\exref{no-chb-path} demonstrates that $\chb{}$ is not the right
relation on events to detect violations of conflict
serializability. Then, what is the right relation to track? In order
to identify that, we will first recast
\defref{conflict-serializability} in terms of events.

We will say that there is a \emph{path} from event $e$ to $f$
\emph{through transactions} in trace $\tr$ (denoted $e \pth{\tr} f$),
if there is a sequence of pairs 
$(e_1,f_1), (e_2,f_2), \ldots
(e_k,f_k)$ ($k > 1$) such that 
\begin{enumerate*}[label=(\alph*)]
\item $e=e_1$ and $f = f_k$,
\item $\trnsc(e_i) = \trnsc(f_i)$, while $\trnsc(f_i) \neq
  \trnsc(e_{i+1})$, for every $i$, and
\item $f_i \chb{\tr} e_{i+1}$ for every $i < k$.
\end{enumerate*}
Using the notion of path between events through transactions, we can
recast the notion of conflict serializability as follows.
\begin{proposition}
\proplabel{conflict-serializability}
A trace $\tr$ is not conflict serializable if and only if there is a
pair of events $e,f$ such that $e \pth{\tr} f$ and $f \chb{\tr} e$.
\end{proposition}


Though $\pth{\tr}$ gives us a characterization of conflict
serializability, it is not clear how to compute it algorithmically in
a single pass over the trace. The reasons are technical and therefore,
skipped. Instead, what we will compute is a slight restriction of the
relation $\pth{\tr}$, defined as follows.
%
%
\begin{definition}
\deflabel{new-edge}
For events $e,f$ in trace $\tr$, we say $e \nedg{\tr} f$, if
there is an event $g$ in $\tr$ such that $e \chb{\tr} g$ and
either 
\begin{enumerate*}[label=(\alph*)]
\item $g = f$, or
\item $g \pth{\tr} f$ and $\trnsc(g)$ is completed in $\tr$.
\end{enumerate*}
\end{definition}

The following theorem formalizes how we can check for
conflict serializability violations using the new relation.
The proof of this theorem is presented in~\cite{aerodrome-tech}.
\begin{theorem}
\thmlabel{nedge-serializability} 
For a transaction $T$, let $T_{\trbegin}$ denote the begin transaction
event $\ev{\cdot,\trbegin}$ of $T$. The following observations hold.
\begin{enumerate}
\item Any trace $\tr$ with a transaction $T$, events $e$ and $f$ such
  that $f \in T$, $e \not\in T$, $T_{\trbegin} \nedg{\tr} e$ and $e
  \nedg{\tr} f$, is not conflict serializable.
\item Let $\tr$ be a trace that is not conflict serializable with a
  witness $T_0,\ldots T_{k-1}$ such that each $T_i$, except possibly
  one, is complete in $\tr$. Then there is a transaction $T$ and
  events $e,f$ in $\tr$ such that $f \in T$, $e \not\in T$,
  $T_{\trbegin} \nedg{\tr} e$ and $e \nedg{\tr} f$.
\end{enumerate}
\end{theorem}

We conclude this section with examples illustrating both the
definition $\nedg{}$ and the use of \thmref{nedge-serializability}.

\begin{example}
\exlabel{nedge-illustration}
Let us begin by looking at trace $\rho_3$ in
\figref{trace-no-chb-path}. Let $\tr_i$ denote the prefix of $\rho_3$
upto (and including) event $e_i$. In trace $\tr_6$, we have $e_3
\nedg{\tr_6} e_6$, $e_4 \nedg{\tr_6} e_5$, and $e_1 \nedg{\tr_6} e_6$
because they are related by $\chb{}$. 
Here, $e_1 \pth{\tr_6} e_4$ because $\trnsc(e_1) = \trnsc(e_3)$,
$e_3 \chb{\tr_6} e_6$ and $\trnsc(e_6) = \trnsc(e_4)$.
However, it is not the case that $e_1 \nedg{\tr_6} e_4$. 
On the other
hand, if we consider $\tr_7$, then $e_1 \nedg{\tr_7} e_4$
as the transaction in $t_1$ is complete in $\tr_7$.
In $\tr_7$
(and therefore also in the full trace $\rho_3$), conditions of
\thmref{nedge-serializability} are satisfied --- $e_1 \nedg{\tr_7}
e_4$ and $e_4 \nedg{\tr_7} e_7$.
\end{example}



\begin{figure}[h]
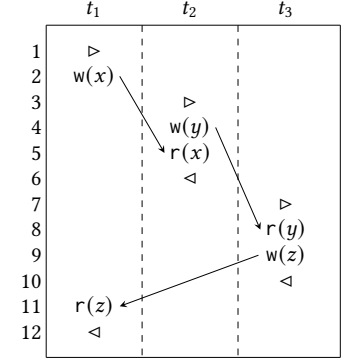

\centering
\execution{3}{
  \figev{1}{\trbeginfig}
  \figev{1}{\wt(x)}
  \figev{2}{\trbeginfig}
  \figev{2}{\wt(y)}
  \figev{2}{\rd(x)}
  \figev{2}{\trendfig}
  \figev{3}{\trbeginfig}
  \figev{3}{\rd(y)}
  \figev{3}{\wt(z)}
  \figev{3}{\trendfig}
  \figev{1}{\rd(z)}
  \figev{1}{\trendfig}
  \orderedge{1}{2}{0.4}{2}{5}{-0.4}
  \orderedge{2}{4}{0.4}{3}{8}{-0.4}
  \orderedge{3}{9}{-0.43}{1}{11}{0.4}
}
\caption{Trace $\rho_4$. Each transaction is a $\thb{}$ predecessor of the other.}
\Description[Trace rho-1 with three threads]{The figure shows a trace, named rho-4, having three threads t-1, t-2 and t-3 performing 12 events in total. 
Event-1 is a begin-transaction event by t-1.
Event-2 is a write event to memory location x by t-1.
Event-3 is a begin-transaction event by t-2.
Event-4 is a write event to memory location y by t-2.
Event-5 is a read event from memory location x by t-2.
Event-6 is an end-transaction event by t-2.
Event-7 is a begin-transaction event by t-3.
Event-8 is a read event from memory location y by t-3.
Event-9 is a write event to memory location z by t-3.
Event-10 is an end-transaction event by t-3.
Event-11 is a read event from memory location z by t-1.
Event-12 is an end-transaction event by t-1.
There is a cross-thread CHB ordering from event-2 (which is a write to memory location x in thread t-1) to event-5 (which is a read from memory location x in thread t-2).
There is a cross-thread CHB ordering from event-4 (which is a write to memory location y in thread t-2) to event-8 (which is a read from memory location y in thread t-3).
There is a cross-thread CHB ordering from event-9 (which is a write to memory location z in thread t-3) to event-11 (which is a read from memory location z in thread t-1).
}
\figlabel{trace-pred-change-cycle}
\end{figure}

\begin{example}
Consider trace $\rho_4$ in \figref{trace-pred-change-cycle}; this is
a slight modification of trace $\rho_1$ from
\figref{trace-pred-change} that now has an atomicity violation. Again
$e_i$ denotes the $i^\text{th}$ event, and $\tr_i$ denotes the prefix upto
event $e_i$. Notice that in prefix $\tr_{11}$, $e_1 \nedg{\tr_{11}}
e_5$ (because $e_1 \chb{\tr_{11}} e_5$) and $e_5 \nedg{\tr_{11}}
e_{11}$ (because $e_5 \pth{\tr_{11}} e_{11}$ 
and $\trnsc(e_5)$ is complete in $\tr_{11}$ ). 
Thus by~\thmref{nedge-serializability}, 
there is a violation of conflict serializability.
\end{example}


\section{Vector Clock Algorithm}
\seclabel{vc}

Based on the intuitions developed in \secref{challenges}, we will now
describe our vector clock based algorithm called {\algo}, for
checking violations of conflict serializability. Before presenting the
algorithm itself, we recall some notation and concepts related to
vector clocks that will be useful.

Let us fix the set of threads in the trace/program to be
$\setofthreads$. A vector time (or timestamp) is a vector of
non-negative integers, whose size/dimension is $|\setofthreads|$
(number of threads).  For a thread $t \in \setofthreads$, we denote
the $t^{\text{th}}$ component of a vector time $V$ by $V(t)$.  We say
a vector time $V_1$ is \emph{less than}
(or \emph{ordered before} or simply \emph{before})
 another time $V_2$ (of the same
dimension), denoted $V_1 \cle V_2$ if $\forall t \in \setofthreads .\:
V_1(t) \leq V_2(t)$.  
In this case, we say that $V_2$ is \emph{greater than, ordered after} or \emph{after} $V_1$.
The minimum vector time on threads
$\setofthreads$ is $\bot_{\setofthreads} = \lambda t.\: 0$, and we
will often use $\bot$ when $\setofthreads$ is clear from context.
Next, the \emph{join} of two vector times $V_1$ and $V_2$ is the time
$V_1 \mx V_2 = \lambda t\cdot \max \set{V_1(t), V_2(t)}$.  Finally, we
use $V[c/t]$ to denote the timestamp $\lambda u.\: \text{ if } u =
t \text{ then } c \text{ else } V(u)$.  Vector clocks are variables
(or place holders) for vector timestamps.  That is, vector clocks are
variables that take values from the space of vector times, and will be
used in our algorithm to compute the timestamps associated with
various events in a trace.  All the operations on vector times can be
naturally thought of as applying to vector clocks as well.


\subsection{The {\algo} Algorithm}
\seclabel{algo}

Our algorithm {\algo} is a single pass linear time
algorithm.
It processes events in
the trace as they are generated and 
(implicitly) assigns vector timestamps to
each of these events. 
Broadly, the goal of the algorithm will
be to assign vector timestamps that capture the relation $\nedg{}$
(\defref{new-edge}) and use \thmref{nedge-serializability} to discover
conflict serializability violations. The exact invariant maintained by
the algorithm is technical and is presented in~\cite{aerodrome-tech}.
Similar to vector clock algorithms used in data race detection
algorithms~\cite{Pozniansky:2003:EOD:966049.781529,fasttrack,wcp2017},
{\algo} does not explicitly store the timestamps of each event
in the trace; 
it instead maintains the timestamps of constantly many events
using constantly many vector clocks.
This small set of vector clocks is adequate 
for detecting conflict serializability violations.


\begin{algorithm*}[t]
\begin{multicols}{2}

\begin{algorithmic}[1]

\Procedure{Initialization}{}
	\For{$t \in \threads{}$}
		\State $\Cc_t$ := $\bot[1/t]$;
		$\Cc^\trbegin_t$ := $\bot$;
	\EndFor
	\For{$\lk \in \locks{}$}
		\State $\Ll_\lk$ := $\bot$;
		$\lastrel_\lk$ := $\nil$;
	\EndFor
	\For{$x \in \vars{}$}
		\State $\Ww_x$ := $\bot$;
		$\lastw_x$ := $\nil$;
		\For{$t \in \threads{}$}
			$\Rr_{t, x}$ := $\bot$;
		\EndFor 
	\EndFor
\EndProcedure
\vspace{0.1in}

\Procedure{checkAndGet}{{\textsf{clk}, \textsf{t}}} \label{line:cag}
	\If{$\Cc^\trbegin_\textsf{t} \cle \textsf{clk}$ and \textsf{t} has an active transaction} \label{line:cag_check}
		\State declare `\textbf{conflict serializability violation}'; \label{line:cag_report}
	\EndIf
	\State $\Cc_\textsf{t}$ := $\Cc_\textsf{t} \mx \textsf{clk}$; \label{line:cag_get}
\EndProcedure
\vspace{0.1in}

\Procedure{acquire}{$t$, $\lk$} %
	\If{$\lastrel_\lk \neq t$} \label{line:acq_lastrel}
		\State \textsc{checkAndGet}($\Ll_\lk$, $t$); \label{line:acq_cag}
	\EndIf
\EndProcedure
\vspace{0.05in}

\Procedure{release}{$t$, $\lk$}
	\State $\Ll_\lk := \Cc_t$; \label{line:rel_set}
	\State $\lastrel_\lk := t$; \label{line:rel_lastrel}
\EndProcedure
\vspace{0.05in}

\Procedure{fork}{$t$, $u$}
   	\State $\Cc_u := \Cc_u \mx \Cc_t$; \label{line:fork_set}
\EndProcedure
\vspace{0.05in}

\Procedure{join}{$t$, $u$}
   	\State \textsc{checkAndGet}($\Cc_u$, $t$); \label{line:fork_cag}
\EndProcedure
\vspace{0.05in}

\Procedure{read}{$t$, $x$}
   	\If{$\lastw_x \neq t$} \label{line:read_lastw}
   		\State \textsc{checkAndGet}($\Ww_x$, $t$); \label{line:read_cag}
   	\EndIf
   	\State $\Rr_{t, x} := \Cc_t$; \label{line:read_set}
\EndProcedure
\vspace{0.05in}

\Procedure{write}{$t$, $x$}
   	\If{$\lastw_x \neq t$} \label{line:write_lastw_check}
   		\State \textsc{checkAndGet}($\Ww_x$, $t$); \label{line:write_w_cag}
   	\EndIf
   	\For{$u \in \threads{}\setminus\set{t}$} \label{line:write_loop}
   		\State \textsc{checkAndGet}($\Rr_{u, x}$, $t$); \label{line:write_r_cag}
   	\EndFor
   	\State $\Ww_x$ := $\Cc_t$; \label{line:write_set}
   	\State $\lastw_x = t$; \label{line:write_lastw_set}
\EndProcedure
\vspace{0.05in}

\Procedure{begin}{$t$} %
	\State $\Cc_t(t)$ := $\Cc_t(t) + 1$ ; \label{line:begin_inc}
	\State $\Cc^\trbegin_t$ := $\Cc_t$ ; \label{line:begin_set}
\EndProcedure
\vspace{0.05in}

\Procedure{end}{$t$}
	\For{$u \in \threads{} \setminus\set{t}$} \label{line:end_thread_handshake_1}
		\If{$\Cc^\trbegin_t \cle \Cc_u$} \label{line:end_thread_handshake_2}
			\State \textsc{checkAndGet}($\Cc_t$, $u$); \label{line:end_thread_handshake_3}
		\EndIf
	\EndFor
	\For{$\lk \in \locks{}$} \label{line:end_lock_handshake_1}
		\State $\Ll_\lk$ := $\Cc^\trbegin_t \cle \Ll_\lk$ ? $\Cc_t \mx \Ll_\lk$ : $\Ll_\lk$; \label{line:end_lock_handshake_2}
	\EndFor
	\For{$x \in \vars{}$} \label{line:end_var_handshake_1}
		\State $\Ww_x$ := $\Cc^\trbegin_t \cle \Ww_x$ ? $\Cc_t \mx \Ww_x$ : $\Ww_x$; \label{line:end_var_handshake_2}
		\For{$u \in \threads{}$} \label{line:end_var_handshake_3}
			\State $\Rr_{u,x}$ := $\Cc^\trbegin_t \cle \Rr_{u, x}$ ? $\Cc_t \mx \Rr_{u, x}$ : $\Rr_{u, x}$; \label{line:end_var_handshake_4}
		\EndFor
	\EndFor
\EndProcedure

\end{algorithmic}
\end{multicols}
\caption{\textit{\algo: Vector Clock Algorithm for Checking Violation of Conflict Serializability}}
\label{algo:update_simple}
\end{algorithm*}

Pseudocode for {\algo} is shown in
Algorithm~\ref{algo:update_simple}. 
It processes events in the trace
based on their operation, calling the appropriate handler. 
As mentioned before, the algorithm uses several vector clocks, 
which we will depict using the black-board font --- $\Cc, \Ll, \Ww, \Rr$, etc. 
Let us assume for now that every event in the trace is part of some
transaction, and that transactions are not nested; later in this
section, we will describe how to efficiently
handle nested transactions and
\emph{unary} transactions, i.e., events not enclosed within a begin
and end atomic block.

\subsubsection{Vector Clocks and Other Data in the State}

The most crucial set of clocks maintained by the algorithm
are those of the form $\Cc_t$, for each thread $t \in \setofthreads$.
The clock $\Cc_t$, intuitively, stores the timestamp of the last event
performed by the thread $t$ so far.
That is, when performing an event $e = \ev{t, op}$,
the timestamp assigned to $e$ by {\algo} is, in fact,
determined by the value of
the clock $\Cc_t$ right after $e$ was processed by the algorithm.
This is similar in spirit to vector clock algorithms
for data race detection such as the standard
\textsc{Djit+}~\cite{Pozniansky:2003:EOD:966049.781529}
or its derivatives like \textsc{FastTrack}~\cite{fasttrack}.
The precise definition of `'the timestamp associated with an event'
is technical and is deferred to~\cite{aerodrome-tech}.

The algorithm also checks for violations of conflict
serializability using the characterization in \thmref{nedge-serializability}, 
which relies on the timestamp of the begin event of a transaction. 
The algorithm, therefore, also maintains
another clock
$\Cc_t^\trbegin$ which intuitively
stores the timestamp of the last begin event
performed by thread $t$.

The goal of these vector timestamps is to capture the relation
$\nedg{}$. Since $\nedg{}$ is defined using $\chb{}$, we need to
ensure that the vector timestamps reflect the orderings induced by $\chb{}$. 
In order to capture the intra-thread dependencies imposed by
$\chb{}$ and $\nedg{}$,
we need auxiliary clocks. 
Consider an event $e$ of the form $\ev{t,\acq(\lk)}$. 
All previously encountered events with operations on lock $\lk$ are
$\chb{}$-before $e$. Hence the timestamp of $e$ must be after those assigned
to such events. To do this, {\algo} will maintain a vector clock
$\Ll_\lk$ for each lock $\lk$, that stores the timestamp of the last
$\rel(\lk)$ seen so far; this will be used to ensure that the
timestamp of $e$ is appropriately larger. Similarly, we need to ensure
that the timestamp of every write event is after the timestamp of all
previous writes and reads to the same variable, and that of a read
event is after the timestamp of previous writes. Therefore, for every
variable $x$, {\algo} has a clock $\Ww_x$ that stores the timestamp of
the last write $\wt(x)$-event and a clock $\Rr_{t,x}$ that stores the
time of the last $\ev{t,\rd(x)}$-event. 

Recall that, when considering
paths between events through transactions ($\pth{}$), we need to make
sure that consecutive transactions along the path are distinct. 
{\algo} tracks this constraint by
maintaining scalar variables $\lastrel_\lk$ and $\lastw_x$, 
which store the identifier of
the thread that performed the last release on $\lk$ and write on $x$,
respectively.

\subsubsection{Initialization and Updates to State}

Each of the clocks $\Cc_t$ are initialized with the time
$\bot[1/t]$, all other clocks are initialized to $\bot$, and all the
scalar variables are initialized to a default value of $\nil$.

As new events are observed in the trace, the algorithm updates these
vector clocks in a manner that is consistent with tracking the
$\nedg{}$-relation.
When processing a begin event $e = \ev{t, \trbegin}$, the algorithm
first increments the \emph{local} component of $\Cc_t$ 
(\cref{line:begin_inc} - `$\Cc_t := \Cc_t[\Cc_t(t) + 1]$'). 
To understand why, let $e_\text{prev}$ 
be some event in the previous transaction
(if any) by the same thread $t$.  
Further, let $e'$ be some event performed by a
different thread $t' \neq t$ such that
\begin{enumerate*}[label=(\alph*)]
\item $e_\text{prev} \nedg{} e'$, and
\item $\neg (e \nedg{} e')$.
\end{enumerate*}
The increment of the local component
ensures that this relationship between 
$e$, $e_\text{prev}$ and $e'$
can be accurately inferred from their timestamps
by ensuring that the local component of the timestamp of $e$
is strictly greater than that of $e_\text{prev}$.  
Finally, {\algo} updates $\Cc_t^\trbegin$ with the timestamp of the current event
$e$ stored in $\Cc_t$.

When processing an acquire event $e = \ev{t, \acq(\lk)}$, the
algorithm makes sure that the timestamp of $e$ is ordered after the
timestamp of the last $\rel(\lk)$-event $e_\lk$ in the trace so far.  
This is achieved by
updating `$\Cc_t := \Cc_t \mx \Ll_\lk$' in the procedure
$\textsc{checkAndGet}$ (invoked at \cref{line:acq_cag}); the procedure
$\textsc{checkAndGet}$ also checks for conflict serializability
violation before updating $\Cc_t$, but more on that later.  Of course,
if $e_\lk$ is performed by the same thread $t$ (\cref{line:acq_lastrel}),
then, this is already ensured and no explicit update is required.

At a write event $e = \ev{t, \wt(x)}$, \algo~ensures that the
timestamp of $e$ is ordered after all the prior reads and writes on
$x$ by calling $\textsc{checkAndGet}$ in \cref{line:write_w_cag,line:write_r_cag}.  The
algorithm then updates $\Ww_x$ to be the timestamp of $e$ 
(see \cref{line:write_set}) and $\lastw_x$ to $t$, 
thus preserving the semantics of the clock
$\Ww_x$ and the scalar variable $\lastw_x$.  The updates performed at
a read event are similar.

At a fork event $e = \ev{t, \fork(u)}$, the algorithm updates the
clock of the child thread $u$ 
(`$\Cc_u := \Cc_u \mx \Cc_t$' in~\cref{line:fork_set}) so
that all events of $u$ are ordered after $e$.  At a join event $e =
\ev{t, \join(u)}$, the algorithm updates $\Cc_t$ to $\Cc_t \mx \Cc_u$
so that all events of thread $u$ are ordered before $e$.

Let us now consider the updates performed at an end-transaction event
$e = \ev{t, \trend}$. Let $e^\trbegin$ denote the matching begin
transaction event. Observe that for an event $f$, if $e^\trbegin
\nedg{} f$, then $e \nedg{} f$ because $\trnsc(e)$ is completed in $\tr$.
That is, all future events that are $\nedg{}$-after $e^\trbegin$ must
be assigned a timestamp after that of $e$. 
This is ensured by updating clocks $\Cc_u$ for all
threads $u$ that satisfy $\Cc_t^\trbegin \cle \Cc_u$ (\cref{line:end_thread_handshake_1,line:end_thread_handshake_2,line:end_thread_handshake_3}), and clocks
$\Ll_\lk$, $\Ww_x$, and $\Rr_{u,x}$ (\cref{line:end_lock_handshake_1,line:end_lock_handshake_2,line:end_var_handshake_1,line:end_var_handshake_2,line:end_var_handshake_3,line:end_var_handshake_4}).
%
%
%

\subsubsection{Checking Violations of Atomicity}

The algorithm detects violations of atomicity at various points by a
call to the procedure \textsc{checkAndGet}. The checks can be broadly
classified into two categories.  First, the algorithm can report a
violation at an event $e = \ev{t, op}$ such that there is an earlier
event $e'$ (performed by a thread $t' \neq t$) that conflicts with
$e$ (and thus $e' \nedg{} e$). 
In this case, if $e^\trbegin \nedg{} e'$ (where $e^\trbegin$ is
the begin event of $\trnsc(e)$), then there is an atomicity
violation as per \thmref{nedge-serializability}.
This check is performed at acquire events (\cref{line:acq_cag}), at read
events (\cref{line:read_cag}) and at write events (\cref{line:write_r_cag,line:write_w_cag}).  
Second, the algorithm reports atomicity violations when
processing an end event $e = \ev{t, \trend}$ (with a matching begin event
$e^\trbegin$). 
The algorithm detects a violation when there is another
thread $u \neq t$ having an active transaction,
with begin event $e^\trbegin_u$ and last event is $e_u$,
such that $e^\trbegin \nedg{} e_u$ (\cref{line:end_thread_handshake_2}) and $e^\trbegin_u \nedg{} e$ (\cref{line:end_thread_handshake_3}). 
These checks for
violations of conflict serializability are performed in
\textsc{checkAndGet} (\cref{line:cag}), which takes two arguments: 
\textsf{clk} (vector timestamp)
and \textsf{t} (thread identifier),
and declares a violation if 
\begin{enumerate*}[label=(\alph*)]
\item thread \textsf{t} has an active
transaction, and 
\item \textsf{clk} is ordered after $\Cc_t^\trbegin$,
which is the timestamp of the
begin event of the (active) transaction of $t$ (\cref{line:cag_check}).
\end{enumerate*} 
Whenever a violation is found, the algorithm exits.
Otherwise, the algorithm continues after updating the
value of the clock $\Cc_\textsf{t}$ to $\Cc_\textsf{t} \mx \textsf{clk}$
(\cref{line:cag_get}).

\subsubsection{Nested and Unary Transactions}

Let us now consider the cases of nested and unary transactions that we
postponed. In the case of nested transactions, it is enough to only
consider the outermost transactions and ignore the inner transactions.
This is because if there is a cycle involving a transaction $T$
that is nested inside another transaction $T'$, then there is clearly
also a cycle involving $T'$.  As a result, we simply ignore the begin
and end events that have a non-zero nesting depth.

Events that are not enclosed by begin and end transaction events
consitute a trivial atomic block, namely, one consisting of only that
single event. These were called \emph{unary} transactions in~\cite{velodrome}. 
Our algorithm does not report a 
violation at unary transactions
(in the procedure \textsc{checkAndGet}, \cref{line:cag_check,line:cag_report}) as
these are not active transactions.
The algorithm, nevertheless, is still correct as
a unary transaction 
(corresponding to a read, write, acquire or join event) 
can only correspond to a cycle that involves another non-unary
transactions. 

We conclude this section with a theorem stating the correctness of
Algorithm~\ref{algo:update_simple} (proof can be found
in~\cite{aerodrome-tech}).

\begin{theorem}
\thmlabel{correctness}  
On any trace $\tr$, Algorithm~\ref{algo:update_simple} reports a
violation of conflict serializability iff $\tr$ is not conflict serializable with a
  witness $T_0,\ldots T_{k-1}$ such that each $T_i$, except possibly
  one, is complete in $\tr$.
\end{theorem}


\subsection{{\algo} on Example Traces}
\seclabel{demo}

Let us illustrate {\algo}'s workings on the traces from
\secref{challenges}. Even though these examples do not use any
synchronization primitives like locking, they contain all the features
needed to highlight the subtle aspects of {\algo}.


\begin{figure}[h]
\begin{minipage}{0.05\textwidth}
\scalebox{1.0}{
\execution{2}{
  \figev{1}{\trbeginfig}
  \figev{2}{\trbeginfig}
  \figev{1}{\wt(x)}
  \figev{2}{\rd(x)}
  \figev{2}{\wt(y)}
  \figev{1}{\rd(y)}
  \figev{1}{\trendfig}
  \figev{2}{\trendfig}
}
}
\end{minipage}
\hspace{1.05in}
\begin{minipage}{0.25\textwidth}
\scalebox{0.8}{
\renewcommand{\arraystretch}{1.13}
\begin{tabular}{cccc}
$\Cc_{t_1}$	& $\Cc_{t_2}$	& $\Ww_x$		& $\Ww_y$	\\ \hline
\vc{2,0}	&				&				&           \\ 
			& \vc{0,2}	&				&           \\ 
			&				& \vc{2,0}	&           \\ 
			& \vc{2,2}	&				&           \\ 
			&				&				& \vc{2,2}\\ 
\multicolumn{4}{l}{\text{Conf. serializ. violation } ($\Cc_{t_1}^\trbegin  \cle \Ww_y$)} \\ 
\\\\
\hline
\end{tabular}
}
\end{minipage}
\caption{{\algo} on Trace $\rho_2$.}
\Description[Run of {\algo} on Trace rho-2 from~\figref{trace-timestamps-events}]
{This figure has two sub-figures. 
On the left, we have the earlier trace rho-2.
We skip describing the sub-figure on the left as it has been described previously in
\figref{trace-timestamps-events}.
On the right, we show the values of the different clocks that {\algo} maintains,
after processing each event in the trace rho-2.
The sub-figure on the right, called "run on rho-2" has a tabular format, with 4 columns and as many rows as the events in rho-2 (i.e., 8 rows).
The first two columns denote the values of the clocks $\Cc_{t-1}$ and $\Cc_{t-2}$.
The next two columns denote the values of the clocks $\Ww_x$ and $\Ww_y$.
We now describe the rows of this subfigure.
The row corresponding to event-1 (which is a begin transaction event by t-1) has the timestamp $\vc{2,0}$ under the column $\Cc_{t-1}$; all other columns are empty.
The row corresponding to event-2 (which is a begin transaction event by t-2) has the timestamp $\vc{0,2}$ under the column $\Cc_{t-2}$; all other columns are empty.
The row corresponding to event-3 (which is a write event to memory location x by t-1) has the timestamp $\vc{2,0}$ under the column $\Ww_{x}$; all other columns are empty.
The row corresponding to event-4 (which is a read event from memory location x by t-2) has the timestamp $\vc{2,2}$ under the column $\Cc_{t-2}$; all other columns are empty.
The row corresponding to event-5 (which is a write event to memory location y by t-2) has the timestamp $\vc{2,2}$ under the column $\Ww_{y}$; all other columns are empty.
The row corresponding to event-6 (which is a read event from memory location y by t-1) has the text "Conflict serializability violation detected" because the clock corresponding to the begin transaction of t-1
is ordered before the clock $\Ww_y$.
The remaining rows are empty.
}
\figlabel{run-timestamps-events}
\end{figure}

Let us begin with the simplest trace $\rho_2$ from
\figref{trace-timestamps-events}.  We show the values of the relevant
vector clocks in~\figref{run-timestamps-events}.  In this figure, we
only depict the value of a vector clock in row $i$ if its value has
changed after processing the $i^\text{th}$ event $e_i$ in the trace. 
We do not show the values of the clocks $\Rr_{t_1, x}$, $\Rr_{t_2, x}$.
$\Rr_{t_1, y}$ or $\Rr_{t_1, y}$ as they are not important here.
There are two threads and thus the size of each vector clock is $2$. 
The clocks $\Cc_{t_1}$ and $\Cc_{t_2}$ are initialized to the
timestamps \vc{1,0} and \vc{0,1} respectively, and all other clocks
are initialized to $\bot = \vc{0,0}$.  The local clocks increment after
a begin event (\cref{line:begin_inc} in Algorithm~\ref{algo:update_simple}) and thus
the clocks $\Cc_{t_1}$ and $\Cc_{t_2}$ become \vc{2,0} and \vc{0,2}
after $e_2$.  Further, these are also the values of the clocks
$\Cc^\trbegin_{t_1}$ and $\Cc^\trbegin_{t_2}$ from this point onwards
until the end of the execution.  After processing $e_3 = \ev{t_1,
  \wt(x)}$, the value of the clock $\Ww_x$ becomes \vc{2,0} (\cref{line:write_set}).
At event $e_4$, the call to \textsc{checkAndGet} (see~\cref{line:read_cag}) with
arguments (\vc{2,0}, $t_2$) updates the clock 
$\Cc_{t_2}$ to \vc{2,2} (\cref{line:cag_get}).  
The clock $\Ww_y$ gets the value of $\Cc_{t_2} = \vc{2,2}$
after processing $e_5$.  
Finally, at event $e_6$, the algorithm calls
\textsc{checkAndGet} with arguments (\vc{2,2}, $t_1$).  In this
procedure, the algorithm asserts that $\Cc_{t_1}^\trbegin \cle \Ww_y$
and declares an atomicity violation.


\begin{figure}[h]
\begin{minipage}{0.05\textwidth}
\scalebox{1.0}{
\execution{2}{
  \figev{1}{\trbeginfig}
  \figev{2}{\trbeginfig}
  \figev{1}{\wt(x)}
  \figev{2}{\wt(y)}
  \figev{1}{\rd(y)}
  \figev{2}{\rd(x)}
  \figev{1}{\trendfig}
  \figev{2}{\trendfig}
}
}
\end{minipage}
\hspace{1.05in}
\begin{minipage}{0.25\textwidth}
\scalebox{0.8}{
\renewcommand{\arraystretch}{1.13}
\begin{tabular}{cccc}
$\Cc_{t_1}$ & $\Cc_{t_2}$ & $\Ww_x$   & $\Ww_y$ \\ \hline
\vc{2,0}    &             &           &         \\
            & \vc{0,2}    &           &         \\
            &             & \vc{2,0}  &         \\
            &             &           & \vc{0,2} \\

\vc{2, 2}   &             &           &         \\
            & \vc{2,2}    &           &         \\
\multicolumn{4}{l}{\text{Conf. serializ. violation } ($\Cc_{t_2}^\trbegin  \cle \Cc_{t_1}$)} \\\\
\hline
\end{tabular}
}
\end{minipage}
\caption{{\algo} on Trace $\rho_3$.}
\Description[Run of {\algo} on Trace rho-3 from~\figref{trace-no-chb-path}]
{This figure has two sub-figures. 
On the left, we have the earlier trace rho-3.
We skip describing the sub-figure on the left as it has been described previously in
\figref{trace-no-chb-path}.
On the right, we show the values of the different clocks that {\algo} maintains,
after processing each event in the trace rho-3.
The sub-figure on the right, called "run on rho-3" has a tabular format, with 4 columns and as many rows as the events in rho-3 (i.e., 8 rows).
The first two columns denote the values of the clocks $\Cc_{t-1}$ and $\Cc_{t-2}$.
The next two columns denote the values of the clocks $\Ww_x$ and $\Ww_y$.
We now describe the rows of this subfigure.
The row corresponding to event-1 (which is a begin transaction event by t-1) has the timestamp $\vc{2,0}$ under the column $\Cc_{t-1}$; all other columns are empty.
The row corresponding to event-2 (which is a begin transaction event by t-2) has the timestamp $\vc{0,2}$ under the column $\Cc_{t-2}$; all other columns are empty.
The row corresponding to event-3 (which is a write event to memory location x by t-1) has the timestamp $\vc{2,0}$ under the column $\Ww_{x}$; all other columns are empty.
The row corresponding to event-4 (which is a write event to memory location y by t-2) has the timestamp $\vc{0,2}$ under the column $\Ww_{y}$; all other columns are empty.
The row corresponding to event-5 (which is a read event from memory location y by t-1) has the timestamp $\vc{2,2}$ under the column $\Cc_{t-1}$; all other columns are empty.
The row corresponding to event-6 (which is a read event from memory location x by t-2) has the timestamp $\vc{2,2}$ under the column $\Cc_{t-2}$; all other columns are empty.
The row corresponding to event-7 (which is an end transaction event by t-1) has the text "Conflict serializability violation detected" because the clock corresponding to the begin transaction of t-2
is ordered before the clock $\Cc_{t-1}$.
The remaining rows are empty.
}
\figlabel{run-no-chb-path}
\end{figure}

Let us next consider the trace $\rho_3$
from~\figref{trace-no-chb-path}. {\algo}'s run on this trace is shown
in \figref{run-no-chb-path}. 
Updates corresponding to the first four events are straightforward.
In event $e_5$, $\Cc_{t_1}$ gets updated to 
\vc{2,2} because of the call to \textsc{checkAndGet} in~\cref{line:read_cag}.  
Notice that this call does not raise
any violation of atomicity because at this point, $\Cc_{t_1}^\trbegin = \vc{2,0}$ 
and the clock $\Ww_y$ is \vc{0,2} thus failing the check
$\Cc_{t_1}^\trbegin \cle \Ww_y$ in~\cref{line:cag_check}.  
The same explanation applies to the $\rd(x)$ event $e_6$ 
in $t_2$ and thus no atomicity violation is reported here as well. 
Next, the algorithm processes the
end event $e_7 = \ev{t_1, \trend}$.  
At this point, the algorithm checks if any event in 
the currently active transaction of $t_2$ 
is ordered after $e_1$ 
(condition $\Cc_{t_1}^\trbegin \cle \Cc_{t_2}$ 
in~\cref{line:end_thread_handshake_2} of Algorithm~\ref{algo:update_simple}).  
This check succeeds since
$\Cc_{t_1}^\trbegin = \vc{2,0}$ and $\Cc_{t_2} = \vc{2,2}$ at this point.
The algorithm then checks if $\Cc_{t_2}^\trbegin \cle \Cc_{t_1}$ in
the procedure \textsc{checkAndGet} and thus declares an atomicity
violation.  This illustrates the subtlety in how
the algorithm reports atomicity violations 
at an end event.


\begin{figure*}[ht]
\centering
\begin{minipage}{0.4\textwidth}
\hspace{0.5in}
\scalebox{1.0}{
\execution{3}{
  \figev{1}{\trbeginfig}
  \figev{1}{\wt(x)}
  \figev{2}{\trbeginfig}
  \figev{2}{\wt(y)}
  \figev{2}{\rd(x)}
  \figev{2}{\trendfig}
  \figev{3}{\trbeginfig}
  \figev{3}{\rd(y)}
  \figev{3}{\wt(z)}
  \figev{3}{\trendfig}
  \figev{1}{\rd(z)}
  \figev{1}{\trendfig}
}
}
\end{minipage}
\begin{minipage}{0.5\textwidth}
\scalebox{0.8}{
\renewcommand{\arraystretch}{1.13}
\begin{tabular}{cccccc}
$\Cc_{t_1}$	& $\Cc_{t_2}$	& $\Cc_{t_3}$ & $\Ww_x$     & $\Ww_y$     & $\Ww_z$ 	    \\ \hline
\vc{2,0,0}  &             &             &             &             &             \\
            &             &             & \vc{2,0,0}  &             &             \\
            & \vc{0,2,0}  &             &             &             &             \\
            &             &             &             & \vc{0,2,0}  &             \\
            & \vc{2,2,0}  &             &             &             &             \\
            &             &             &             & \vc{2,2,0}  &             \\
            &             & \vc{0,0,2}  &             &             &             \\
            &             & \vc{2,2,2}  &             &             &             \\
            &             &             &             &             & \vc{2,2,2}  \\
\\
\multicolumn{6}{c}{\text{Conf. serializ. violation } ($\Cc_{t_1}^\trbegin  \cle \Ww_z$)} \\
\\
\hline
\end{tabular}
}
\end{minipage}
\caption{{\algo} on Trace $\rho_4$.}
\Description[Run of {\algo} on Trace rho-4 from~\figref{trace-pred-change-cycle}]
{This figure has two sub-figures. 
On the left, we have the earlier trace rho-4.
We skip describing the sub-figure on the left as it has been described previously in
\figref{trace-pred-change-cycle}.
On the right, we show the values of the different clocks that {\algo} maintains,
after processing each event in the trace rho-4.
The sub-figure on the right, called "run on rho-4" has a tabular format, with 6 columns and as many rows as the events in rho-4 (i.e., 12 rows).
The first three columns denote the values of the clocks $\Cc_{t-1}, \Cc_{t-2}$ and $\Cc_{t-3}$.
The next three columns denote the values of the clocks $\Ww_x, \Ww_y$ and $\Ww_z$.
We now describe the rows of this subfigure.
The row corresponding to event-1 (which is a begin transaction event by t-1) has the timestamp $\vc{2,0,0}$ under the column $\Cc_{t-1}$; all other columns are empty.
The row corresponding to event-2 (which is a write event to memory location x by t-1) has the timestamp $\vc{2,0,0}$ under the column $\Ww_{x}$; all other columns are empty.
The row corresponding to event-3 (which is a begin transaction event by t-2) has the timestamp $\vc{0,2,0}$ under the column $\Cc_{t-2}$; all other columns are empty.
The row corresponding to event-4 (which is a write event to memory location y by t-2) has the timestamp $\vc{0,2,0}$ under the column $\Ww_{y}$; all other columns are empty.
The row corresponding to event-5 (which is a read event from memory location x by t-2) has the timestamp $\vc{2,2,0}$ under the column $\Cc_{t-2}$; all other columns are empty.
The row corresponding to event-6 (which is a end transaction event by t-2) has the timestamp $\vc{2,2,0}$ under the column $\Ww_{y}$; all other columns are empty.
The row corresponding to event-7 (which is a begin transaction event by t-3) has the timestamp $\vc{0,0,2}$ under the column $\Cc_{t-3}$; all other columns are empty.
The row corresponding to event-8 (which is a read event from memory location x by t-3) has the timestamp $\vc{2,2,2}$ under the column $\Cc_{t-3}$; all other columns are empty.
The row corresponding to event-9 (which is a write event to memory location z by t-3) has the timestamp $\vc{2,2,2}$ under the column $\Ww_{z}$; all other columns are empty.
The row corresponding to event-10 (which is an end transaction event by t-3) is completely empty.
The row corresponding to event-11 (which is a read event from memory location z by t-1) has the text "Conflict serializability violation detected" because the clock corresponding to the begin transaction of t-1 is ordered before the clock $\Ww_{z}$.
The remaining rows are empty.
}
\figlabel{run-pred-change}
\end{figure*}

We will now illustrate how Algorithm~\ref{algo:update_simple} detects
the atomicity violation in the more involved trace $\rho_4$ from
\figref{trace-pred-change-cycle}. This example illustrates how {\algo}
handles dependencies between transactions introduced by future events.
The run of {\algo} on $\rho_4$ is shown in \figref{run-pred-change}.
We omit the updates to the clocks $\Rr_{t_i, u}$ ($i \in \set{1, 2,
  3}$, $u \in \set{x, y, z}$) as they do not play a significant role
in this example.  All vector clocks have dimension $3$ because there
are three threads in $\rho_4$.  As before, the clocks are initialized
as follows: $\Cc_{t_1} = \vc{1,0,0}$, $\Cc_{t_2} = \vc{0,1,0}$ and
$\Cc_{t_3} = \vc{0,0,1}$; all other clocks are initialized to
\vc{0,0,0}.  The begin events result in incrementing of
local clocks and thus $\Cc_{t_1} = \vc{2,0,0}$ after $e_1$.  Further,
the clock $\Ww_x$ gets updated to the value of $\Cc_{t_1}$ at the end
of $e_2$.  The next two events $e_3$ and $e_4$ are processed in a
similar fashion.  
At event $e_5 = \ev{t_2, \rd(x)}$, the clock
$\Cc_{t_2}$ gets updated to \vc{2,2,0} 
(\cref{line:cag_get} in Algorithm~\ref{algo:update_simple}).  
After this, the
transaction in $t_2$ ends. The clocks of none of the threads is
updated because of $e_6$ as neither thread $t_1$ nor $t_3$ have clock
values larger than $\Cc_{t_2}^\trbegin$ (\cref{line:end_thread_handshake_2}). 
However the write and read clocks are updated. 
Specifically, the clock $\Ww_y$
maintaining the timestamp to the last write to $y$ is such that
$\Cc_{t_2}^\trbegin\cle \Ww_y$ and thus, the algorithm updates $\Ww_y$
to $\Ww_y \mx \Cc_{t_2} = \vc{2,2,0}$ 
(\cref{line:end_var_handshake_2} in Algorithm~\ref{algo:update_simple}).  
Event $e_7$ is a begin event and
updates $\Cc_{t_3}$ to \vc{0,0,2}.  Now at the $\rd(y)$ event $e_8$,
the clock $\Cc_{t_3}$ gets updated with $\Ww_y$ which at this point
evaluates to \vc{2,2,0}, thus giving $\Cc_{t_3} = \vc{2,2,2}$.  The
write clock $\Ww_z$ then gets updated to \vc{2,2,2} after $e_9$.
More clock updates happen at $e_{10}$
(though not shown in~\figref{run-pred-change})
Finally, an atomicity violation is detected at
event $e_{11} = \ev{t_1, \rd(z)}$; the algorithm checks if the clock
$\Ww_z$ \emph{knows} some event in $t_1$ ($\Cc_{t_1}^\trbegin \cle
\Ww_z$) and declares a violation of conflict
serializability as this check passes.


\subsection{Reducing the number of Read Clocks}
\seclabel{read_clock_opt}

Recall that Algorithm~\ref{algo:update_simple} maintains, a vector
clock $\Rr_{t,x}$ for every pair of thread $t$ and memory location
$x$.  Therefore, the number of such vector clocks that need to be tracked
in the basic algorithm is $O(|\setofthreads|V)$, where $|\setofthreads|$ is the number of threads
and $V$ is the number of memory locations.
Storing and updating these many clocks can be expensive,
when the number of memory locations that need to be tracked
is prohibitively large, as is the case for most real world software.
We tackle this using our optimization to reduce the number
of clocks from $O(|\setofthreads|V)$ to $O(V)$.  
To understand the optimization, we need to first understand
the role served by clocks $\Rr_{t,x}$. 
First, these clocks help detect
atomicity violation --- at a write event $e = \ev{t, \wt(x)}$, 
a violation is reported if there is a thread $u \neq t$ such that
$\Cc^\trbegin_t \cle \Rr_{u, x}$ 
(\cref{line:cag_check} invoked from~\cref{line:write_r_cag} in Algorithm~\ref{algo:update_simple}).  
Second, these clocks are used to
update $\Cc_t$ --- at a write event $e = \ev{t, \wt(x)}$, we set
$\Cc_t := \bigsqcup_{u \neq t} \Cc_t \mx \Rr_{u, x}$ 
(\cref{line:cag_get} invoked iteratively at~\cref{line:write_r_cag}).

The reduction in the number of clocks is achieved by instead maintaining
one clock (per memory location) 
for each of the above two purposes
instead of maintaining $O(|\setofthreads|)$ many clocks (per memory location).
First, for updating clocks correctly at write events,
we will maintain a single clock $\Rr_x$ for each
location $x$.
This clock stores the value $\bigsqcup_u
\Rr_{u,x}$ at each point while processing the trace. 
Next, to perform checks for violations of conflict
serializability, we will have another clock $\chR_x$ (\emph{check read}). 
This clock will store the value $\bigsqcup_u \Rr_{u,x}[0/u]$ 
at each point in the analysis. 
Based on the invariants maintained by
the algorithm, one can show that checking $\Cc^\trbegin_t \cle
\bigsqcup_{u\neq t} \Rr_{u,x}$ is equivalent to checking
$\Cc^\trbegin_t \cle \chR_x$.
This optimization and other useful optimizations that improve the
performance of~\algo, are outlined in greater detail
in~\cite{aerodrome-tech}.

We now state the time and space complexity for the optimized version
discussed in this section. We will use $n_\textsf{non-end}$ and
$n_\textsf{end}$ for the number of non-end events and end events in the
trace (and thus $n = n_\textsf{non-end} + n_\textsf{end}$ is the
size of the trace).  We will denote by $|\setofthreads|$, $V$ and $L$ the number of
threads, memory locations and locks in the input trace.  Further, all
arithmetic operations are assumed to take constant time. 
\begin{theorem}
\thmlabel{complexity}
The algorithm takes $O(|\setofthreads|(n_\textsf{non-end} + (|\setofthreads|+L+V)n_\textsf{end}))$
time and $O(|\setofthreads|(|\setofthreads|+V+L))$ space.
\end{theorem}

The complexity observations easily follow from the description of the
algorithm and the optimization discussed in~\secref{read_clock_opt}.


\section{Experimental Evaluation}
\seclabel{experiments}

In this section, we describe our implementation of~{\algo}
and the results of evaluating it on benchmark programs.
\appref{artifact} discusses the accompanied artifact that
describes our overall experimental workflow and 
can be used to replicate our results.

\subsection{Implementation}

We have implemented {\algo} in a prototype tool {\tool},
available publicly~\cite{rapid}.
{\tool} is written in Java and analyzes traces generated 
by concurrent programs to detect violations of conflict
serializability.  
The primary goal of the evaluation is to assess if the theoretical
bound (linear time) of the algorithm also translates to
effective performance in practice, or in other words,
\emph{does our vector clock algorithm perform better than existing 
approaches  such as the classical graph based algorithm 
(Velodrome) proposed in~\cite{velodrome}?}
We emphasize that the primary purpose of the evaluation is
to compare different \emph{algorithms} for checking atomicity
instead of comparing different \emph{tools} that implement these algorithms.

\subsubsection*{Logging.}
In order to evaluate our algorithm against the above objective
and to ensure a fair comparison with other approaches, we must
ensure that all competing candidate algorithms analyze the same trace.
However, the dynamic behavior of a concurrent program
can vary significantly across different runs, 
even when starting with the same input.
In order to ensure fairness, we compare the performance of
the different algorithms on the \emph{same} dynamic execution.
Our tool {\tool} therefore first extracts an 
execution trace from a concurrent programs
and then analyzes the same trace against all candidate algorithms.
We use RoadRunner~\cite{flanagan2010roadrunner} to
log traces from our set of benchmark programs. 
RoadRunner uses load time program instrumentation and can be
extended to log various events --- read and write accesses to memory locations, 
acquire and release of synchronization objects (locks),
forks and joins of threads, and
events generated at the entry and exit of each method, 
which we respectively mark as transaction begin ($\trbegin$) and end ($\trend$) events.

\subsubsection*{Velodrome.}
The Velodrome algorithm~\cite{velodrome} runs in (worst case) cubic time and
analyzes traces by building a directed graph, with
transactions as nodes in the graph and where the edges correspond to
$\thb{}$ relation between transactions.
There was no publicly available implementation of Velodrome that analyzes logged executions.
Thus, we also implement this algorithm in {\tool}.  
We use the Java graph library JGraphT~\cite{jgrapht} 
to implement various graph operations (adding nodes and edges,
cycle detection, etc.,) in Velodrome algorithm.
In our implementation of Velodrome, we
also incorporate garbage collection as an optimization suggested
in~\cite{velodrome} --- transactions with no incoming edges do not
participate in cycles and can be deleted from the graph.  
In line with the objective of our evaluation,
we analyze {\algo} and Velodrome on the same trace 
(generated by RoadRunner) to ensure a fair comparison.

\subsubsection*{Other techniques.}
The tool DoubleChecker~\cite{doublechecker} is a state-of-the-art tool
for checking conflict serializability in a sound and complete manner.
DoubleChecker implements a two-phase analysis --- the first
phase performs a fast but imprecise analysis and reports 
an over-approximation of the actual set of cycles in the transaction graph.
The second phase then filters out the false positives from this set
with a more fine grained analysis.
DoubleChecker's performance crucially relies on
the first phase being carried out while the program executes.
Therefore, one cannot get performance data for 
DoubleChecker on a logged trace.
As a result, there can be no fair comparison between our algorithm
and DoubleChecker as one cannot guarantee that the two analyses 
run on the same trace.
In order to gauge if DoubleChecker will significantly outperform
our implementation of {\algo}, we ran DoubleChcker's publicly
available implementation~\cite{doublechecker-tool}
on a subset of our benchmarks. 
On these benchmarks, DoubleChecker's performance was slower by an order
of magnitude. 
While these experiments do not indicate that DoubleChecker performs
worse than our algorithm,
they do suggest that
our algorithm will be competitive against DoubleChecker.
We choose not to present these numbers in this paper,
because they are not an apples-to-apples comparison.


\subsection{Atomicity Specifications and Benchmarks}

\subsubsection*{Atomicity Specifications.}
\seclabel{atom_spec}
In general, the logging mechanism in RoadRunner instruments and tracks
all events corresponding to entering and exiting methods.  A na\"{i}ve
atomicity specification would be to mark all method boundaries as
atomic.  However, as expected, not all methods are intended to be
atomic.  For example, default methods like \texttt{run} or
the static \texttt{main} methods in Java are often not intended to be atomic.
Thus, atomicity specifications need to be specially identified by
developers, by supplying manual annotations~\cite{fq03}.  In the
absence of such static annotations, we use atomicity specifications
from prior work~\cite{doublechecker} whenever possible
(Table~\ref{tab:time2}).
For the benchmarks (Table~\ref{tab:time1}) for which no specifications were available,
we declare all methods except the \texttt{main} and \texttt{run}
methods to be atomic.



\begin{table*}[t]
\caption{
Trace characteristics and running times for benchmarks with atomicity specifications from DoubleChecker.
}
\vspace{-0.1in}
\centering
\scalebox{0.85}{
\begin{adjustbox}{center}
\renewcommand{\arraystretch}{1.2}
\begin{tabular*}{2.4043\columnwidth}{!{\VRule[1pt]}c!{\VRule[1pt]}c|c|c|c|c!{\VRule[1pt]}c|c|c|c!{\VRule[1pt]}}
\specialrule{1pt}{0pt}{0pt}
1 & 2 & 3 & 4 & 5 & 6 & 7 & 8 & 9 & 10\\ 
\specialrule{1pt}{0pt}{0pt}
\cellcolor[HTML]{EFEFEF} \quad Program \quad\,
& \cellcolor[HTML]{EFEFEF} \, Events \,
& \cellcolor[HTML]{EFEFEF} \, Threads \,
& \cellcolor[HTML]{EFEFEF} \, Locks \,
& \cellcolor[HTML]{EFEFEF} Variables
& \cellcolor[HTML]{EFEFEF} Transactions 
& \cellcolor[HTML]{EFEFEF} Atomic? 
& \cellcolor[HTML]{EFEFEF} Velodrome (s)
& \cellcolor[HTML]{EFEFEF} \algo~(s)
& \cellcolor[HTML]{EFEFEF} Speed-up\\
\specialrule{1pt}{0pt}{0pt}
\textsf{avrora}			& 2.4B	& 7 	& 7 	& 1079K	& 498M	& \xmark 	& TO 	& 1.5 & $>24000$ \\
\textsf{elevator}		& 280K	& 5		& 50	& 725	& 22.6K & \cmark	& 162	& 1.7 &	$97$	\\
\textsf{hedc}			& 9.8K 	& 7		& 13	& 1694	& 84	& \xmark	& 0.07	& 0.06 	& $1.16$ \\
\textsf{luindex}			& 570M 	& 3		& 65 	& 2.5M 	& 86M 	& \xmark	& 581 	& 674 & $0.86$ \\
\textsf{lusearch}		& 2.0B 	& 14	& 772 	& 38M 	& 306M 	& \xmark	& TO 	& 5.5 & $>6545$ \\
\textsf{moldyn}			& 1.7B 	& 4 	& 1 	& 121K 	& 1.4M 	& \xmark	& TO 	& 54.9 & $>650$ \\
\textsf{montecarlo}		& 494M 	& 4 	& 1 	& 30.5M & 812K 	& \xmark	& TO 	& 0.75 & $>48000$ \\
\textsf{philo}			& 613	& 6		& 1		& 24	& 0		& \cmark	& 0.02	& 0.02	& $1$ \\
\textsf{pmd}				& 367M 	& 13 	& 223 	& 12.9M & 81M 	& \xmark	& 3.1 	& 3.8 & $0.82$ \\
\textsf{raytracer}		& 2.8B 	& 4 	& 1 	& 12.6M & 277M 	& \cmark	& TO	& 55m40s & $>10.7$ \\
\textsf{sor}				& 608M 	& 4 	& 2 	& 1M 	& 637K 	& \xmark	& 6.9 	& 9.6 & $0.72$ \\
\textsf{sunflow}			& 16.8M & 16 	& 9 	& 1.2M 	& 2.5M 	& \xmark	& 67.9 	& 0.65 & $104.5$ \\
\textsf{tsp} 			& 312M 	& 9		& 2		& 181M	& 9		& \xmark	& 4.2	& 5.7	& $0.73$	\\
\textsf{xalan}			& 1.0B	& 13 	& 8624 	& 31M 	& 214M 	& \xmark	& 1.6 	& 2.0 & $0.8$ \\
\specialrule{1pt}{0pt}{0pt}
\end{tabular*}
\end{adjustbox}
}
\label{tab:time2}
\end{table*}




\begin{table*}[t]
\caption{
Trace characteristics and running times for benchmarks with naive atomicity specifications.
}
\vspace{-0.1in}
\centering
\scalebox{0.85}{
\begin{adjustbox}{center}
\renewcommand{\arraystretch}{1.2}
\begin{tabular*}{2.4034\columnwidth}{!{\VRule[1pt]}c!{\VRule[1pt]}c|c|c|c|c!{\VRule[1pt]}c|c|c|c!{\VRule[1pt]}}
\specialrule{1pt}{0pt}{0pt}
1 & 2 & 3 & 4 & 5 & 6 & 7 & 8 & 9 & 10\\ 
\specialrule{1pt}{0pt}{0pt}
\cellcolor[HTML]{EFEFEF} Program
& \cellcolor[HTML]{EFEFEF} \, Events \,
& \cellcolor[HTML]{EFEFEF} \, Threads \,
& \cellcolor[HTML]{EFEFEF} \, Locks \,
& \cellcolor[HTML]{EFEFEF} Variables
& \cellcolor[HTML]{EFEFEF} Transactions 
& \cellcolor[HTML]{EFEFEF} Atomic? 
& \cellcolor[HTML]{EFEFEF} Velodrome (s)
& \cellcolor[HTML]{EFEFEF} \algo~(s)
& \cellcolor[HTML]{EFEFEF} Speed-up\\
\specialrule{1pt}{0pt}{0pt}
\textsf{batik}			& 186M 	& 7 	& 1916	& 4.9M 	& 15M 	& \xmark	& 52.7 & 65.5 & $0.81$ \\
\textsf{crypt} 			& 126M 	& 7 	& 1 	& 9M 	& 50 	& \xmark	& 92.1 & 104 & $0.88$ \\
\textsf{fop} 			& 96M 	& 1 	& 115 	& 5M 	& 25M 	& \cmark	& 88.3 & 92.5 & $0.95$ \\
\textsf{lufact} 			& 135M 	& 4 	& 1 	& 252K 	& 642M	& \xmark	& 2.4 & 2.9 & $0.82$ \\
\textsf{series} 			& 40M 	& 4 	& 1 	& 20K 	& 20M 	& \xmark	& 61.0 & 15.3 & $3.98$ \\
\textsf{sparsematmult}	& 726M	& 4 	& 1 	& 1.6M 	& 25 	& \xmark	& 1210 & 1197 & $1.01$ \\
\textsf{tomcat}			& 726M	& 4 	& 1 	& 1.6M 	& 25 	& \xmark	& 3.4 & 4.5 & $0.75$ \\
\specialrule{1pt}{0pt}{0pt}
\end{tabular*}
\end{adjustbox}
}
\label{tab:time1}
\end{table*}

\subsubsection*{Benchmarks and Setup.}
\seclabel{benchmarks}
Our benchmark programs (Table~\ref{tab:time2} and
Table~\ref{tab:time1}) are derived from the DaCaPo benchmark
suite~\cite{DaCapo2006} adapted to run with 
RoadRunner~\cite{flanagan2010roadrunner}, 
Java Grande Forum~\cite{JGF2001} and microbenchmarks from~\cite{vonPraun03}
and have been used in prior work~\cite{doublechecker}.  
Our experiments were conducted on a 2.6GHz 64-bit Linux machine with Java
1.8 as the JVM and 30GB heap space.  In each table, Column 1 depicts
the name of the benchmark.  Column 2 reports the number of events in
the trace generated from the corresponding benchmark program in Column
1.  Observe that the number of events in the execution traces
can vary from a few hundred to billions of events
and our algorithm can scale to such large traces.
Column 3, 4 and 5 report the number of distinct threads, locks and
variables accessed in the trace generated.  Column 6 reports the
number of transactions in the trace.  Column 7 reports `\xmark' if an
atomicity violation was detected and reports `\cmark' otherwise.
Columns 8 and 9 report the time (in seconds) taken by respectively the
Velodrome algorithm and {\algo} introduced in this article to analyze
the trace generated; a `TO' represents timeout after 10
hours.  Column 10 reports the speed-up of {\algo} over
Velodrome.

\subsection{Evaluation Results}

For the first set of benchmarks (Table~\ref{tab:time2}), 
we use the atomicity specification obtained from prior work~\cite{doublechecker}.
For the second set of benchmarks (Table~\ref{tab:time1}),
we use default atomicity specifications (all methods
except \texttt{main} and \texttt{run} are assumed to be atomic).
The specifications from~\cite{doublechecker} are carefully
crafted to ensure that spurious atomicity violations are not reported.
In the absence of careful specifications, we can expect
that the violations will be reported early on in executions.

Let us first consider the first set of benchmarks from
Table~\ref{tab:time2}.  On most of these benchmarks, the violations of
atomicity are discovered late in the trace.  This is expected as the
specifications are realistic and do not declare all methods to be
atomic.  The performance of~\algo~is significantly better than that of
Velodrome.  Velodrome times out on most of these benchmarks (time
limit was set to be 10 hours).  This is because of the prohibitively
large number of transactions that get accumulated in these traces.
Consider, for example, the case of \textsf{sunflow} for which~\algo~takes less
than a second, while Velodrome spends about $68$ seconds.  
In this benchmark, the number of nodes in the graph analyzed by Velodrome is about
$9000$, at the point where the violation is reported. 
This coupled with the cubic runtime complexity, results in
the notable slowdown.  Notice that, the slowdown is despite the
garbage collection optimization implemented in Velodrome.  Our
algorithm, on the other hand, has a linear running time.  
Similarly, in the benchmark \textsf{avrora}, the number of
transactions is more than $393$K 
in the prefix of the trace in which {\algo}
reports an atomicity violation.
Any super linear time analysis
is unlikely to scale for so many transactions, and Velodrome, in fact,
does not return an answer within $10$ hours.  {\algo}, on the other
hand, scales to traces with more than a billion events (\textsf{avrora},
\textsf{lusearch}, \textsf{moldyn}, \textsf{raytracer}, \textsf{xalan}) and demonstrates the effectiveness
of a linear time vector clock algorithm.
For the examples on which {\algo} does not give a huge speedup over
Velodrome, we discovered that the number of
nodes in Velodrome's graph analysis is fairly small
owing to garbage collection;
for example, there were 13 nodes in the graph for \textsf{pmd}, 4
nodes in \textsf{sor} and 13 nodes in \textsf{xalan}.

In the second set of benchmarks, we notice that the performance of
Velodrome is comparable to that of our algorithm~\algo.
This is expected because the atomicity specifications are 
inadequate and do not reflect realistic ones --- typically
most methods are non-atomic and developers have
to identify a smaller set of candidate code blocks that
they think are atomic.
As a result, on these benchmarks, violations are detected early on
in the trace and thus, the size of the transaction graph
in Velodrome's analysis is small.
A detailed analysis of the traces suggests that in all these benchmarks,
the number of nodes in the transaction graph constructed by Velodrome did not grow more than $4$, except for
\textsf{tomcat}, for which the size of the graph grows to $21$.
In this case, the cost of maintaining vector clocks and updating
them at every event overrides their potential benefits,
and as a result, the graph based algorithm runs faster.


\section{Related Work}
\seclabel{related}

Multi-threaded programs are challenging to reason about.
Atomicity is a principled concept that lets programmers
reason about coarse behaviors of programs, without
being concerned about fine grained thread interleavings.
Ensuring atomicity of concurrent program blocks is therefore
an important question~\cite{lpsz08} and has been investigated thoroughly.

Static analysis techniques analyze source code to confirm
the atomicity of code blocks marked atomic.
Such techniques prominently rely on the design of type systems~\cite{Flanagantypesatomicity2008,fq03}.
These type systems rely on commutativity of operations
and are inspired from Lipton's theory of reduction~\cite{Lipton:1975:RMP:361227.361234}
and the concept of purity~\cite{ffq05}. 
Extensions to type inference~\cite{saws05} 
and to programs with non-blocking synchronization~\cite{ws05} 
have been developed.
The work in~\cite{Flanagantypesatomicity2008} uses constraint based 
type system inference for inferring atomicity specifications.

Dynamic analysis algorithms for checking atomicity inspect individual program executions
instead of the program source code.
Lipton's theory of reduction~\cite{Lipton:1975:RMP:361227.361234} has been
a prominent theme in this space, most notably the analysis employed by Atomizer~\cite{ff04}.
This approach however leads to false alarms.
The notion of conflict serializability was introduced concurrently
by Flanagan et. al.~\cite{velodrome} and Farzan et. al.~\cite{fm08}, 
inspired from the theory
of concurrency control in databases~\cite{Papadimitriou:1986:TDC:6174}.
However, Farzan et. al.~\cite{fm08} do not account for 
any lock operations which are crucially used in most 
Java like concurrent programs, making their algorithm
prone to false positives.
Further, their algorithm relies on maintaining
sets of locks, threads and variables, similar in spirit
to the Goldilocks algorithm~\cite{elmas2007goldilocks} 
for detecting HB races.
As in the case of data race detection~\cite{fasttrack,Kini2018},
such an algorithm is expected to be orders of magnitude 
slower than a vector clock algorithm for the same problem. 
More importantly, the algorithm in~\cite{fm08} is automata-theoretic,
warranting a global centralized observer that analyzes events in a
serial fashion. In contrast, our algorithm {\algo}
allows for a distributed implementation --- one can attach the 
analysis metadata (vector clocks and other scalar variables, in our case) 
to the various objects (like threads, locks and memory locations) 
being tracked.
The analysis can then be performed with only little synchronization between these metadata, allowing our vector clock algorithm to leverage parallelism.
Recently, DoubleChecker~\cite{doublechecker} proposed a two-pass analysis
for efficient detection of conflict serializability violations.
Here, a coarse first pass detects potential cycles in the
transaction graph. This is followed by a fine grained analysis
that tracks more information and ensures the soundness of the overall analysis.
Causal atomicity~\cite{fm06} is a weaker
criterion for atomicity and
asks if there is an equivalent trace where one particular transaction
(instead of all transactions) is serial.

As with most concurrency bugs, detecting atomicity violations is a
challenging problem and is subject to interleaving explosion problem.
Techniques such as that in CTrigger~\cite{ctrigger2009} and
AVIO~\cite{ltqz06} resort to directed exploration of thread
interleavings to expose subtle atomicity violations.
Penelope~\cite{penelope2010} detects 2 thread atomicity violations
using directed interleaving exploration.  The work
in~\cite{wangatomicity2006a,wangatomicity2006b,Lu2012,Agarwal:2005:ORR:1101908.1101944}
is also based on exercising specific thread schedules.  SMT solving
based predictive analysis
techniques~\cite{Wang:2010:TSA:2175554.2175589} have been developed,
but tend to not scale.  The work of Samak
et. al.~\cite{Samak:2015:STD:2786805.2786874} synthesizes directed
unit tests for catching atomicity violations.  The work
in~\cite{fm06,Sen:2006:MCM:2135909.2135950} develop techniques for
model checking concurrent programs for exposing atomicity violations.
The use of random sampling and thread scheduling have also been
proposed previously in the
literature~\cite{Joshi:2009:CEA:1575060.1575118,Park:2008:RAA:1453101.1453121}.

Like most concurrency bugs, atomicity bugs are hard to 
fix. Naive fixes such as enforcing atomic regions using locks can 
introduce new bugs, affect the performance of programs 
and moreover can be inadequate in ensuring atomicity. 
Several approaches have been proposed~\cite{Jin2011,Liu2016,Lin2018,Li2019,Liu2012,Lin2014,Li2019}
for automated repair of atomicity violation bugs.

\section{Conclusions}
\seclabel{conclusions}

In this paper, we considered the problem of checking atomicity in
concurrent programs.  Conflict serializability of traces is a popular
notion for checking atomicity dynamically.  We present the first
linear time, vector clock algorithm for checking violations of
conflict serializability on traces of concurrent programs.  Our
experimental evaluation demonstrates the power of a linear time
algorithm, in that, it scales well to large executions and is often
faster than existing graph based algorithms.  Interesting avenues for
future work include extending the insights developed in our paper to
design efficient algorithms for other notions of atomicity, including
causal atomicity~\cite{fm06}, view
serializability~\cite{wangatomicity2006b} or reduction based atomicity
characterizations as in~\cite{ff04,wangatomicity2006a}.  Other
promising lines of work include improving the efficiency of the
proposed dynamic analysis for atomicity by incorporating ideas from
data race detection.  This includes the classic \emph{epoch}
optimizations~\cite{fasttrack}, static analysis for redundancy
elimination~\cite{redcard} and optimal check placement~\cite{bigfoot},
and advances concerning instrumentation~\cite{Bond2013,Wilcox2015,Cao2016,Wood2017}.

\begin{acks}
We thank the anonymous reviewers for several comments that helped improve the paper.
Umang Mathur is partially supported by a Google PhD Fellowship. 
Mahesh Viswanathan is partially supported by NSF CCF 1901069.
\end{acks}

\bibliographystyle{ACM-Reference-Format}
\bibliography{references}
\clearpage

\appendix


\section{Proof of for~\thmref{nedge-serializability}}
\applabel{app-proof}

\begin{proof}
$(\Leftarrow)$ Observe that for any pair of events 
$e_1,e_2$ if $\trnsc(e_1) \neq \trnsc(e_2)$ and $e_1
  \nedg{\tr} e_2$ then $e_1 \pth{\tr} e_2$. 
This means $T_\trbegin \pth{\tr} e \pth{\tr} f$.
Since $\trnsc(f) = \trnsc(T_\trbegin)$, we can rewrite this as
$T_\trbegin \pth{\tr} e \pth{\tr} T_\trbegin$ or simply
$T_\trbegin \pth{\tr} T_\trbegin$.
The rest of the proof follows from \propref{conflict-serializability},
the observation that $T_\trbegin \chb{\tr} T_\trbegin$.

$(\Rightarrow)$ 
Let $T_0, \ldots, T_{k-1}$ 
be a witness sequence for the conflict serializability
violation of $\tr$
($k>1$ and $T_i \neq T_j$ for every $i\neq j$).
Then, we must have a sequence of pairs of events $(e_0,f_0), \ldots
(e_{k-1},f_{k-1})$ such that $\trnsc(e_i) = \trnsc(f_i) = T_i$,
and $f_i \chb{\tr} e_{(i+1)\bmod k}$. 
Observe that for every $i \neq j$, we have $e_i \pth{\tr} f_j$.
Let $m$ be the index of the only active transaction in $\tr$
amongst $\set{T_i}_{i=0}^{k-1}$;
if all transactions are completed, pick $m=0$).
Now let $T = T_m$,
$e = e_{(m+1)\bmod k} \not\in T$ and $f = e_m \in T$.
Now, $T_\trbegin \chb{\tr} f_m \chb{\tr} e_{(m+1)\bmod k} = e'$
and thus $T_\trbegin\nedg{\tr} e'$.
Also, because of the choice of $m$, the transaction
$\trnsc(e)$ is completed in $\tr$ and $e \pth{\tr} f$
and thus $e \nedg{\tr} f$.
\end{proof}


\section{Correctness of {\algo}}
\applabel{correctness}

\newcommand{\evnt}[2]{\mathsf{ev}^{#1}_{#2}}

We now prove that Algorithm~\ref{algo:update_simple} reports a
violation on a trace $\tr$ if and only if $\tr$ is not conflict
serializable (as per \defref{conflict-serializability}). The key is to
identify the invariant being maintained by the algorithm. Intuitively,
the vector clocks track the $\nedg{}$ dependencies, but the precise
invariant is technical. We need to introduce some notation to state it
precisely.

Consider a complete observed trace $\tr$. For an event $e \in \tr$,
$\mathsf{prefix}^\tr(e)$ is the shortest prefix of $\tr$ that contains
$e$. For an arbitrary prefix $\pi$ of $\tr$, we will find it useful to
introduce notation for identifying some specific events in $\pi$. For a pair $p
= \ev{t,op}$, $\evnt{\pi}{p}$ denotes the last event of the form $p$
in $\pi$; note that for some pairs $p$, 
this maybe undefined as there might be no event of this form in $\pi$. 
Thus, for example, $\evnt{\pi}{\ev{t,\trbegin}}$ denotes the last
transaction begin event performed by thread $t$ in $\pi$. Sometimes,
it will be convenient to leave one of the two arguments in the pair $p
= \ev{t,op}$ unspecified, and in this case $\evnt{\pi}{p}$ will denote
the last event of type identified by the specified argument. Thus, for
example, $\evnt{\pi}{\ev{\cdot,\wt(x)}}$ is the last $\wt(x)$-event in
$\pi$ (regardless of the thread performing it), and
$\evnt{\pi}{\ev{t,\cdot}}$ is the last event of thread $t$ in $\pi$
(regardless of the operation). For an event $e$, let us define $B(e)$
to be the number of $\ev{\thread(e),\trbegin}$ events in
$\mathsf{prefix}^\tr(e)$, i.e., $B(e)$ is the number of begin
transaction events performed by $\thread(e)$ before $e$ (including $e$). Finally, to
state the invariant, we identify the timestamp of an event in the
prefix. This timestamp changes as we process more of the trace. The
vector timestamp $C(e, \pi)$ of event $e$ in prefix $\pi$ is given by
\[
C(e,\pi)(u) = \left\{ \begin{array}{l}
              B(e)+1, \qquad \mbox{if } u = \thread(e)\\
              \max \{B(f)+1\: |\: f = \ev{u,op} \mbox{ and }
                          f \nedg{\pi} e\}, \\
               \qquad \mbox{otherwise}
                     \end{array} \right.
\]
We can now state the invariant that identifies the values of all the
vector clocks maintained by the algorithm.

\begin{lemma}
\lemlabel{invariant}
After any prefix $\pi$ of $\tr$, Algorithm~\ref{algo:update_simple}
stores the following values.
\[
\begin{array}{ll}
\Cc_t = C(\evnt{\pi}{\ev{t,\cdot}},\pi) &
  \Cc_t^\trbegin = C(\evnt{\pi}{\ev{t,\trbegin}},\mathsf{prefix}^\tr(\evnt{\pi}{\ev{t,\trbegin}})) \\
  \Rr_{t,x} = C(\evnt{\pi}{\ev{t,\rd(x)}},\pi) &
\Ww_x = C(\evnt{\pi}{\ev{\cdot,\wt(x)}},\pi) \\
  \Ll_\lk = C(\evnt{\pi}{\ev{\cdot,\rel(\lk)}},\pi)
\end{array}
\]
\end{lemma}

The lemma is proved by an induction on the length of the trace
processed by the algorithm. The proof is straightforward, and skipped.
The invariant allows us to establish the correctness of the
algorithm. The proof of \thmref{correctness} follows from
\thmref{nedge-serializability} and \lemref{invariant}.


\section{Optimizations for~\algo}
\applabel{app-opt}

\subsection{Read Clocks}

\begin{algorithm*}[ht]
\begin{multicols}{2}

\begin{algorithmic}[1]

\Procedure{Initialization}{}
	\For{$t \in \threads{}$}
		\State $\Cc_t$ := $\bot[1/t]$;
		$\Cc^\trbegin_t$ := $\bot$;
	\EndFor
	\For{$\lk \in \locks{}$}
		\State $\Ll_\lk$ := $\bot$;
		$\lastrel_\lk$ := $\nil$;
	\EndFor
	\For{$x \in \vars{}$}
		\State $\Ww_x$ := $\bot$;
		$\lastw_x$ := $\nil$;
		\State $\Rr_x$ := $\bot$;
		$\chR_x$ := $\bot$; 
	\EndFor
\EndProcedure
\vspace{0.1in}

\Procedure{checkAndGet}{{\textsf{clk}$_1$, \textsf{clk}$_2$, \textsf{t}}} %
	\If{$\Cc^\trbegin_\textsf{t} \cle \textsf{clk}_1$ and \textsf{t} has an active transaction}
		\State declare `\textbf{conflict serializability violation}';
	\EndIf
	\State $\Cc_\textsf{t}$ := $\Cc_\textsf{t} \mx \textsf{clk}_2$;
\EndProcedure
\vspace{0.1in}

\Procedure{read}{$t$, $x$}
   	\If{$\lastw_x \neq t$}
   		\State \textsc{checkAndGet}($\Ww_x$, $\Ww_x$, $t$);
   	\EndIf
   	\State $\Rr_x := \Rr_x \mx \Cc_t$;
   	\State $\chR_x := \chR_x \mx \Cc_t[0/t]$;
\EndProcedure
\vspace{1in}

\Procedure{write}{$t$, $x$}
   	\If{$\lastw_x \neq t$}
   		\State \textsc{checkAndGet}($\Ww_x, \Ww_x$, $t$);
   	\EndIf
   	\State \textsc{checkAndGet}($\chR_x, \Rr_x$, $t$);
   	\State $\Ww_x$ := $\Cc_t$;
   	\State $\lastw_x = t$;
\EndProcedure
\vspace{0.05in}

\Procedure{end}{$t$}
	\For{$u \in \threads{} \setminus\set{t}$}
		\If{$\Cc^\trbegin_t \cle \Cc_u$}
			\State \textsc{checkAndGet}($\Cc_t$, $\Cc_t$, $u$);
		\EndIf
	\EndFor
	\For{$\lk \in \locks{}$}
		\State $\Ll_\lk$ := $\Cc^\trbegin_t \cle \Ll_\lk$ ? $\Cc_t \mx \Ll_\lk$ : $\Ll_\lk$;
	\EndFor
	\For{$x \in \vars{}$}
		\State $\Ww_x$ := $\Cc^\trbegin_t \cle \Ww_x$ ? $\Cc_t \mx \Ww_x$ : $\Ww_x$;
		\If{$\Cc^\trbegin_t \cle \Rr_x$}
			\State $\Rr_x$ := $\Cc_t \mx \Rr_x$;
			\State $\chR_x$ := $\Cc_t[0/t] \mx \chR_x$;
		\EndIf
	\EndFor
\EndProcedure

\end{algorithmic}
\end{multicols}
\caption{\textit{\algo: Reducing the number of read clocks. Only procedures that differ from Algorithm~\ref{algo:update_simple} have been presented.}}
\label{algo:update_read_clock_opt}
\end{algorithm*}

The algorithm maintains the invariant that
for two events $e_1$ and $e_2$ with $\thread(e_1) = t_1$, we have 
$C_{e_1} \cle C_{e_2}$ iff $C_{e_1}(t_1) \leq C_{e_2}(t_1)$.
In other words, in order to compare the timestamps of two
events, it is enough to compare the \emph{local time}
corresponding to the thread of the smaller timestamp.
Now, observe that, at a write event $e = \ev{t, \wt(x)}$, the algorithm
detects an atomicity violation by either comparing with
the clock of the last write event ($\Ww_x$), or by 
comparing with the clocks of the last read events of
each thread, except the thread $t$.
Let us consider the second check.
Observe that in this case, a violation is raised
if there is a thread $u \neq t$ such that $\Cc_t^\trbegin \cle \Rr_{u, x}$.
Based on our earlier observation about local times,
this check is equivalent to the check 
$\exists u\neq t \cdot \Cc_t^\trbegin(t) \cle \Rr_{u, x}(t)$.
Now observe that
\begin{align*}
\begin{array}{rcl}
\exists u\neq t \cdot \Cc_t^\trbegin(t) \cle \Rr_{u, x}(t) &
\text{iff} &
\Cc_t^\trbegin(t) \cle \bigsqcup\limits_{u \neq t} \Rr_{u, x}(t) \\
& \text{iff} & \\
\Cc_t^\trbegin(t) \cle \bigsqcup\limits_{u} \Rr_{u, x}[0/u](t) &
\text{iff} &
\Cc_t^\trbegin \cle \bigsqcup\limits_{u} \Rr_{u, x}[0/u]
\end{array}
\end{align*}
Based on the above observation, we can perform the check
for atomicity if we have a single clock that maintains the timestamp
$\bigsqcup_{e_{t,\rd(x)}} C_{e_{t,\rd(x)}}[0/t]$,
where $e_{u,\rd(x)}$ 
is the last event of the form $\ev{u, \rd(x)}$
seen in the trace so far.
For this, will use a new single clock $\chR_x$ to store
this value and inductively maintain this in the algorithm.

Next, observe that the algorithm ensures that
the timestamp of the last event in a given thread is
larger than the timestamp of any earlier event in the same thread.
This means that, at any point, $\Rr_{x, t} \cle \Cc_t$
and thus we have $\Rr_x \mx \Cc_t = \Cc_t$ at any point in the algorithm.
Now, let us consider how the algorithm updates
$\Cc_t$ with the various $\Rr_{u, x}$ clocks at a write event.
Precisely, if an atomicity violation is not detected
when comparing with the read clocks, the value of
$\Cc_t$ becomes $C^\text{old}_t \mx \bigsqcup\limits_{u \neq t} \Rr_{u, x}$,
where $C^\text{old}_t$ is the value of $\Cc_t$ before the updates.
Coupled with our previous observation, this new value is the
same as the value 
$C^\text{old}_t \mx \Rr_{t, x} \mx \bigsqcup\limits_{u \neq t} \Rr_{u, x}$
which, in turn, can be re-written as 
$C^\text{old}_t \mx \bigsqcup\limits_{u} \Rr_{u, x}$.
Thus, we can maintain the timestamp $\bigsqcup_{e_{t,\rd(x)}} C_{e_{t,\rd(x)}}$
in a single clock ($e_{u,\rd(x)}$ as before,
is the last event of the form $\ev{u, \rd(x)}$
seen in the trace so far).
We use a new clock $\Rr_x$ to maintain this value.

We present the read clock optimization in Algorithm~\ref{algo:update_read_clock_opt}.


\subsection{Other Optimizations}
\seclabel{opt}

We now discuss some additional optimizations
that help improve the runtime performance and memory overhead
of~\algo. These are presented in Algorithm~\ref{algo:update_opt}.

\paragraph*{Lazy Clock Updates}
This optimization is based on the following observation.
Many times, a given memory location $x$ is repeatedly 
read from in by a single thread, before being written to.
This means that the algorithm updates the clocks 
$\Rr_x$ and $\chR_x$ (or the clocks $\Rr_{t, x}$ 
in line 26 of Algorithm~\ref{algo:update_simple})
repeatedly, a lot of times, without being 
used to compute other clocks (lines 31 and 46 
in Algorithm~\ref{algo:update_simple}) or to detect atomicity violation.
When the length of such contiguous subsequence 
of reads is large, these updates to $\Rr_x$ and $\chR_x$
are often redundant.
To cater for this, we update the
$\Rr_x$ clocks in a lazy fashion as follows.  
For every memory location, we maintain a set $\stale^r_x$,
which is the set of threads $t$ that
have performed a read on $x$ after the last
write to $x$ in the current transaction of $t$. 
And then, at a write event $e =\ev{t, \wt(x)}$,
we use the values of the clocks $\setpred{\Cc_u}{u \in \stale^r_x}$ 
to update $\Cc_t$, $\Rr_x$ and $\chR_x$.
This optimization therefore allows us to avoid expensive vector clock
operations at (the majority of) read events
in lieu of cheaper set operations (adding a thread to $\stale^r_x$).
An analogous optimization also applies for the $\Ww_x$ clocks.

\paragraph*{Maintaining Sets of Memory Locations to be updated}
Notice that at an end event (line 43 in
Algorithm~\ref{algo:update_simple}), we check, for every memory
location $x$, whether two clocks are ordered, and if so, perform clock
updates accordingly.  The set of memory locations in the entire trace
can however be prohibitively large, and comparing vector clocks can be
expensive (when performed for every location at every end event).  We
observed that most of the times, memory locations are often local to a
small set of threads, and thus often, clock comparisons in line 43 are
often redundant.  We optimize the number of comparisons by
maintaining, for every thread $t$, the set of memory locations that
have a read or write event ordered after some event in the (unique)
active transaction of $t$.  Then, at an end event, we only need to
iterate over this potentially smaller subset of memory locations.

\paragraph*{Garbage Collection}
This optimization is inspired from the garbage collection
mechanism described in~\cite{velodrome}.
The basic idea there is the following.
If a transaction $T$ is such that there is no event $e$ in the transaction
that is ordered (using $\chb{}$) after some event
of another transaction, then $T$ cannot participate in any cycle
and the analysis can essentially ignore such a transaction.
This optimization can easily be implemented using vector clocks
as follows. In order to check if a transaction of thread $t$
has an incoming edge, we check if either the transaction that
forked $t$ is active or if there is a $u \neq t$
such that $\Cc^\trbegin_t(u) \neq \Cc_t(u)$ at the end of the transaction.







\begin{algorithm*}[ht]
\begin{multicols}{2}

\begin{algorithmic}[1]

\Procedure{Initialization}{}
	\For{$t \in \threads{}$}
		\State $\Cc_t$ := $\bot[1/t]$;
		$\Cc^\trbegin_t$ := $\bot$;
		\State $\upset^r_t$ := $\emptyset$;
		$\upset^w_t$ := $\emptyset$;
	\EndFor
	\For{$\lk \in \locks{}$}
		\State $\Ll_\lk$ := $\bot$;
		$\lastrel_\lk$ := $\nil$;
	\EndFor
	\For{$x \in \vars{}$}
		\State $\Ww_x$ := $\bot$;
		$\lastw_x$ := $\nil$;
		\State $\Rr_x$ := $\bot$;
		$\chR_x$ := $\bot$;
		\State $\stale^r_x$ := $\emptyset$;
		$\stale^w_x$ := $\nil$;
	\EndFor
\EndProcedure
\vspace{0.2in}

\Procedure{hasIncomingEdge}{$t$} %
	\State \Return ($\parenttr_t$ is alive) $\lor$ ($\Cc^\trbegin_t[0/t]\neq \Cc_t[0/t]$);
\EndProcedure
\vspace{0.1in}

\Procedure{checkAndGet}{{\textsf{clk}$_1$, \textsf{clk}$_2$, \textsf{t}}} %
	\If{$\Cc^\trbegin_\textsf{t} \cle \textsf{clk}_1$ and \textsf{t} has an active transaction}
		\State declare `\textbf{conflict serializability violation}';
	\EndIf
	\State $\Cc_\textsf{t}$ := $\Cc_\textsf{t} \mx \textsf{clk}_2$;
\EndProcedure
\vspace{0.1in}

\Procedure{acquire}{$t$, $\lk$} %
	\If{$\lastrel_\lk \neq t$}
		\State \textsc{checkAndGet}($\Ll_\lk, \Ll_\lk$, $t$);
	\EndIf
\EndProcedure
\vspace{0.05in}

\Procedure{release}{$t$, $\lk$}
	\State $\Ll_\lk := \Cc_t$;
	\State $\lastrel_\lk := t$;
\EndProcedure
\vspace{0.05in}

\Procedure{fork}{$t$, $u$}
   	\State $\Cc_u := \Cc_u \mx \Cc_t$;
\EndProcedure
\vspace{0.05in}

\Procedure{join}{$t$, $u$}
   	\State \textsc{checkAndGet}($\Cc_u, \Cc_u$, $t$);
\EndProcedure
\vspace{0.05in}

\Procedure{read}{$t$, $x$}
   	\If{$\lastw_x \neq t$}
   		\If{$\stale^w_x = \true$}
   			\State \textsc{checkAndGet}($\Cc_{\lastw_x}, \Cc_{\lastw_x}$, $t$);
   		\Else
   			\State \textsc{checkAndGet}($\Ww_x$, $\Ww_x$, $t$);
   		\EndIf
   	\EndIf
   	\State $\stale^r_x$ := $\stale^r_x \cup \set{t}$;
   	\For{$u \in \threads{}$}
   		\If{$u$ has an active transaction and $\Cc^\trbegin_u \cle \Cc_t$}
   			\State $\upset^r_u$ := $\upset^r_u \cup \set{x}$;
   		\EndIf
   	\EndFor
\EndProcedure
\vspace{0.05in}

\Procedure{write}{$t$, $x$}
   	\If{$\lastw_x \neq t$}
   		\If{$\stale^w_x = \true$}
   			\State \textsc{checkAndGet}($\Cc_{\lastw_x}, \Cc_{\lastw_x}$, $t$);
   		\Else
   			\State \textsc{checkAndGet}($\Ww_x, \Ww_x$, $t$);
   		\EndIf
   	\EndIf
	\For{$u \in \stale^r_x$}
   		\State $\Rr_x$ := $\Rr_x \mx \Cc_u$;
   		\State $\chR_x$ := $\chR_x \mx \Cc_u[0/u]$;
   	\EndFor
   	\State $\stale^r_x$ := $\emptyset$
   	\State \textsc{checkAndGet}($\chR_x, \Rr_x, t$);
	\State $\stale^w_x := \true$;
   	\State $\lastw_x = t$;
   	\For{$u \in \threads{}$}
   		\If{$u$ has an active transaction and $\Cc^\trbegin_u \cle \Cc_t$ }
   			\State $\upset^w_u$ := $\upset^w_u \cup \set{x}$;
   		\EndIf
   	\EndFor
\EndProcedure
\vspace{0.05in}

\Procedure{begin}{$t$} %
	\State $\Cc_t(t)$ := $\Cc_t(t) + 1$ ;
	\State $\Cc^\trbegin_t$ := $\Cc_t$ ;
\EndProcedure
\vspace{0.05in}

\Procedure{end}{$t$}
	\If{\textsc{hasIncomingEdge}$(t)$}
		\For{$u \in \threads\setminus\set{t}$}
			\If{$\Cc^\trbegin_t \cle \Cc_u$}
				\State \textsc{checkAndGetClock}($\Cc_t$, $\Cc_t$, $u$);
			\EndIf
		\EndFor
		\For{$\lk \in \locks{}$}
			\State $\Ll_\lk$ := $\Cc^\trbegin_t \cle \Ll_\lk$ ? $\Cc_t \mx \Ll_\lk$ : $\Ll_\lk$;
		\EndFor
		\For{$x \in \upset^w_t$}
			\If{$\stale^w_x = \false \lor \lastw_x = t$}
				\State $\Ww_x$ := $\Cc_t \mx \Ww_x$;
			\EndIf
			\If{$\lastw_x = t$}
				\State $\stale^w_x$ := $\false$;
			\EndIf
		\EndFor
		\State $\upset^w_t$ := $\emptyset$;
		\For{$x \in \upset^r_t$}
			\State $\Rr_x$ := $\Cc_t \mx \Rr_x$;
			\State $\chR_x$ := $\chR_x \mx \Cc_t[0/t]$;
			\State $\stale^r_x$ := $\stale^r_x \setminus \set{t};$
		\EndFor
		\State $\upset^r_t$ := $\emptyset$; \\
	\Else
		\For{$x \in \upset^r_t$}
			\State $\stale^r_x$ := $\stale^r_x \setminus \set{t};$
		\EndFor
		\State $\upset^r_t$ := $\emptyset$;
		\For{$x \in \upset^w_t$}
			\If{$\lastw_x = t$}
				\State $\stale^w_x$ := $\false$;
				\State $\lastw_x := \nil$;
			\EndIf
		\EndFor
		\State $\upset^w_t$ := $\emptyset$;
		\For{$\lk \in \locks{}$}
			\If{$\lastrel_\lk = t$}
				\State $\lastrel_\lk := \nil$;
			\EndIf
		\EndFor
	\EndIf
\EndProcedure

\end{algorithmic}
\end{multicols}
\caption{\textit{Optimized version of~\algo}}
\label{algo:update_opt}
\end{algorithm*}

\clearpage




%
%
%
%
%

\section{Artifact Appendix}
\applabel{artifact}

\subsection{Abstract}

This artifact appendix describes how to replicate
our results from~\secref{experiments}. 
Our evaluation comprises of generating trace logs of 
benchmark programs from Table~\ref{tab:time2} and Table~\ref{tab:time1}, 
and running {\algo} and Velodrome~\cite{velodrome} analyses on them.
We expect the speed-ups of {\algo} over Velodrome to be similar to 
those reported in Table~\ref{tab:time2} and Table~\ref{tab:time1}.
All analyses are implemented in our tool {\tool} and we provide
Python scripts for automating the workflow.



\subsection{Artifact check-list (meta-information)}

{\small
\begin{itemize}
  \item {\bf Algorithm: } {\algo}.

  \item {\bf Program: }
  Provided with the artifact (also see~\secref{benchmarks}). 
  
  \item {\bf Data set: }
  Instructions and scripts for generating trace logs from benchmarks programs have been provided. 
  Trace logs used in our original experiments can be downloaded from~\cite{traces}.

  \item {\bf Execution: } Experiments to be conducted as sole user.
  Generating trace logs from scratch can take several hours for large benchmarks.
  

  \item {\bf How much disk space required (approximately)?: } 
  Approximately 500GB space required to save trace logs.
  Individual trace logs can be as large as 100GB.
  
  \item {\bf How much time is needed to prepare workflow (approximately)?: } All scripts are provided.

  \item {\bf How much time is needed to complete experiments (approximately)?: } If all traces need to be generated, then about a day. If traces are obtained from~\cite{traces}, then as much as the timeout set. 
  We used a timeout of 10 hours per benchmark.

  \item {\bf Publicly available?: } Yes. 
  Artifact available at~\cite{aeZenodo}.
  {\tool} available at~\cite{rapid} (archived at~\cite{rapidZenodo}).

  \item {\bf Code licenses (if publicly available)?: } MIT License.

  \item {\bf Data licenses (if publicly available)?: } None.

  \item {\bf Archived (provide DOI)?: } Yes~\cite{aeZenodo}.
\end{itemize}

\subsection{Description}

\subsubsection{How to access} 
Publicly available~\cite{aeZenodo}.
It extracts to less than 250MB.

\subsubsection{Hardware dependencies}
No special hardware required.

\subsubsection{Software dependencies}
Java 1.8 or higher, Ant 1.10 or higher, Python 2.7 or higher.

\subsubsection{Data sets}
Traces can be generated using benchmark programs provided.
Alternatively, they can be downloaded from~\cite{traces}.

\subsection{Installation}
Obtain the artifact from~\cite{aeZenodo} and extract.



\newcommand{\aehome}{\texttt{\$AE\_HOME}}
\newcommand{\fulltr}{\texttt{full\_trace.rr}}
\renewcommand{\tr}{\texttt{trace.std}}

\subsection{Experiment workflow}
\applabel{workflow}

\subsubsection{Directory Structure}

The overall directory of the artifact is shown in~\figref{dir}.
The directory \texttt{benchmarks/} contains our benchmark programs.
The directory \texttt{atomicity\_specs/} contain atomicity specifications for each benchmark (see~\secref{atom_spec}).
The directory \texttt{scripts/} contains our scripts for automating the workflow.
The directory \texttt{RoadRunner} has been obtained from~\cite{flanagan2010roadrunner}.
\texttt{README.md} is a more verbose description of the experimental workflow,
and \texttt{LICENSE.txt} is an MIT License agreement for the artifact.

\begin{figure}[!ht]
\begin{verbatim}
AE/
|--- LICENSE.txt
|--- README.md
|--- RoadRunner/
|--- atomicity_specs/
|--- benchmarks/
|--- scripts/
\end{verbatim}
\caption{Directory structure of the artifact}
\Description{Structure of the directory \texttt{AE/}. It has 2 files, names "LICENSE.txt" and "README.md" and 4 sub-directories named "RoadRunner", "atomicity_specs", "benchmarks" and "scripts".}
\figlabel{dir}
\end{figure}

\subsubsection{Overall Workflow}

The overall workflow is as follows.

\begin{enumerate}
	\item \textbf{Generating Trace Logs}. We need to generate trace logs from benchmark programs. There are two options here:
	\begin{enumerate}[label=(\alph*)]
		\item \textbf{Option-1.} Download trace logs directly from~\cite{traces}.
		\item \textbf{Option-2} (time consuming). Use RoadRunner to generate raw trace logs and then filter those based on the provided atomicity specifications, described below.
			\begin{enumerate}[label=(\roman*)]
				\item \textbf{Logging.} 
				We will use the logging and instrumentation facility provided by RoadRunner~\cite{flanagan2010roadrunner} to generate traces.
				\item \textbf{Filtering.} We will filter out some events based on atomicity specifications in \texttt{atomicity\_specs/}.
			\end{enumerate}
	\end{enumerate}
	\item \textbf{Performing Analyses}  We then analyze the final trace logs (obtained int he previous step) using {\tool}~\cite{rapid}.
	{\tool} can perform several kinds of analyses on a trace log:
	\begin{itemize} 
	\item The class \texttt{MetaInfo} can be used to determine basic information about the log, including the total number of events, threads, variables, locks etc.
	\item The class \texttt{Aerodrome} determines atomicity violations using our proposed algorithm Aerodrome.
	\item The class \texttt{Velodrome} determines atomicity violations using the prior state-of-the-art algorithm Velodrome~\cite{velodrome}.
\end{itemize}
\end{enumerate}

\subsubsection{Getting Started}

Downloaded the artifact from~\cite{aeZenodo} and set {\aehome}:
\begin{verbatim}
> export AE_HOME=/path/to/AE/
\end{verbatim}
Also, you need to change the variable \texttt{home} in the file \texttt{scripts/util.py} (line 17) to be the value of~\aehome .
Also set the environment variables \texttt{JAVA\_HOME} and \texttt{JVM\_ARGS} in the same file appropriately.
Next download {\tool} from GitHub~\cite{rapid} or from the archive~\cite{rapidZenodo}
in \texttt{\$AE\_HOME/rapid/} and install:
\begin{verbatim}
> cd $AE_HOME/rapid/; ant jar
\end{verbatim}

\subsubsection{Generating Trace Logs}
\noindent\\
\textbf{Option-1.}
Readers interested in simply reproducing the results can download the traces used in our experiments from ~\cite{traces}
and jump to~\appref{perfanalyses} directly. 
Next, replace the \texttt{benchmarks/} folder: 
\begin{verbatim}
> rm -rf $AE_HOME/benchmarks/ 
> unzip /path/to/downloaded/zip -d $AE_HOME/ 
> mv $AE_HOME/asplos20-ae-traces $AE_HOME/benchmarks/
\end{verbatim}

\noindent\\
\textbf{Option-2.}
\begin{enumerate}
	\item \textbf{Download and install Roadrunner} 
	\begin{verbatim}
> cd $AE_HOME
> git clone git@github.com:stephenfreund\
/RoadRunner.git
> wget https://raw.githubusercontent.com/umangm/\
rapid/master/notes/PrintSubsetTool.java.txt \
-O $AE_HOME/RoadRunner/src/rr/simple/\
PrintSubsetTool.java
> cd $AE_HOME/RoadRunner; ant; source msetup
\end{verbatim}

\item \textbf{Download and install {\tool}} 
\begin{verbatim}
> cd $AE_HOME
> git clone git@github.com:umangm/rapid.git
> cd $AE_HOME/rapid; ant jar
\end{verbatim}

\item \textbf{Move to \texttt{scripts/} folder.}
We will now execute some scripts and for this, we will change the working directory:
\begin{verbatim}
> cd $AE_HOME/scripts/
\end{verbatim}

\item \textbf{Extract execution logs.}
To generate full trace for a single benchmark:
\begin{verbatim}
> python gen_trace.py <b>
\end{verbatim}
Here, \texttt{<b>} could be something like \texttt{philo}.
To generate traces for all benchmarks:
\begin{verbatim}
> python gen_trace.py
\end{verbatim}
This step generates files {\fulltr} in the directory
\texttt{\$AE\_HOME/benchmarks/<b>/}, either for particular benchmark or for all benchmarks based on the command.

\item \textbf{Atomicity specifications.}
To modify the trace to account for the atomicity specifications for a single benchmark:
\begin{verbatim}
> python atom_spec.py <b>
\end{verbatim}
To account for the atomicity specifications for all benchmarks:
\begin{verbatim}
> python atom_spec.py
\end{verbatim}
This step generates files \texttt{\$AE\_HOME/benchmarks/<b>/trace.std} (either for particular benchmark or for all benchmarks).
At this point, the files \texttt{full\_trace.rr} can be deleted.
\end{enumerate}

\subsubsection{Performing Analyses}
\applabel{perfanalyses}

Our experiments perform 3 different analysis on the traces:
\begin{enumerate*}[label = (\alph*)]
	\item metadata analysis to collect information about the different kinds of events in traces,
	\item Velodrome analysis, and
	\item {\algo} analysis.
\end{enumerate*} \\

\noindent
\textbf{Obtaining Trace metadata.}
To generate metadata information from the trace	of a single benchmark \texttt{<b>}:
\begin{verbatim}
> python metainfo.py <b>
\end{verbatim}
When the files \texttt{trace.std} are available for all benchmarks, run:
\begin{verbatim}
> python metainfo.py
\end{verbatim}
This step generates three files in
\texttt{\$AE\_HOME/benchmarks/<b>/}:
\begin{enumerate*}[label=(\roman*)]
	\item the file \texttt{metainfo.err} contains error information from the Java commands run in the python script \texttt{metainfo.py} and
	should ideally be empty,
	\item \texttt{metainfo.txt} contains the actual output (including the number of different kinds of events);
	refer to \texttt{\$AE\_HOME/README.md} for a description of the contents of this file,
	\item \texttt{metainfo.tim} reports the time taken by the system.
\end{enumerate*}
\\

\noindent
\textbf{{\algo} analysis.}
For a single benchmark \texttt{<b>}, run:
\begin{verbatim}
> python aerodrome.py <b>
\end{verbatim}
%
To analyze the traces for all benchmarks, run:
\begin{verbatim}
> python aerodrome.py
\end{verbatim}
%
This step generates three files in \texttt{\$AE\_HOME/benchmarks/<b>/} - 
\texttt{aerodrome.txt}, \texttt{aerodrome.err} and \texttt{aerodrome.txt}.
Their description can be found in \texttt{\$AE\_HOME/README.md}.
\\

\noindent
\textbf{Velodrome analysis}
For a single benchmark \texttt{<b>}, run:
\begin{verbatim}
> python velodrome.py <b>
\end{verbatim}
To analyze the traces of all benchmarks, run:
\begin{verbatim}
> python velodrome.py
\end{verbatim}
As before this step generates files \texttt{velodrome.txt},
\texttt{velodrome.err} and \texttt{velodrome.tim}, whose description can
be found in the readme file \texttt{\$AE\_HOME/README.md}.

%

\subsection{Evaluation and expected result}

The workflow described in~\appref{workflow} can be used to generate
the data showed in Table~\ref{tab:time2} and Table~\ref{tab:time1}.
The primary objective of the evaluation is to measure the speedup of
{\algo} analysis over Velodrome analysis.
We expect that Aerodrome outperforms Velodrome on all 
benchmarks where the speedup of {\algo} (over Velodrome) is more than $10\times$.
The exact speed-ups may vary depending upon the hardware and other processes running, but orders of magnitude (for speedup) should stay the same.
Of course, results can vary when the the traces used are different from those
used in our experiments~\cite{traces}.
The metadata analysis described in~\appref{workflow}
can be used to generate the total number of events, threads, locks
and memory locations (often referred to as variables).

\subsection{Experiment customization}

All the different analysis described in the workflow (\appref{workflow})
can be performed for an execution of any concurrent Java program.
For this, see the instructions\footnote{\texttt{\$AE\_HOME/rapid/blob/master/notes/Generate\_RoadRunner\_traces.md}} in {\tool}~\cite{rapid,rapidZenodo}
to generate a trace from a benchmark.
After this, if an atomicity specification is available, 
one can account for it by using the script \texttt{atom\_spec.py}.
If not, simply use an empty file for an atomicity specification
and use the same script.
Finally, all the three analyses can be run using scripts provided
(\texttt{metainfo.py}, \texttt{aerodrome.py} and \texttt{velodrome.py}).

\subsection{Notes}

Contact \href{mailto:umathur3@illinois.edu}{\nolinkurl{umathur3@illinois.edu}} regarding any questions.

\subsection{Methodology}

Submission, reviewing and badging methodology:

\begin{itemize}
  \item \url{http://cTuning.org/ae/submission-20190109.html}
  \item \url{http://cTuning.org/ae/reviewing-20190109.html}
  \item \url{https://www.acm.org/publications/policies/artifact-review-badging}
\end{itemize}

\end{document}